\documentclass[a4paper,superscriptaddress,amsmath,amssymb,11pt]{article}

\usepackage{epjcpub} 

\usepackage[T1]{fontenc} 
\usepackage[utf8]{inputenc}
\usepackage{graphicx}
\usepackage{dcolumn}
\usepackage{bm}
\usepackage{float}
\usepackage{multirow}
\usepackage{verbatim}
\usepackage{amsmath}
\usepackage{graphicx,subfigure,epsfig}
\usepackage{mathrsfs}
\usepackage{hyperref}
\hypersetup{
    colorlinks=true,
    citecolor=blue,
    linkcolor=blue,
    filecolor=magenta,
    urlcolor=black,}
\usepackage{color}

\newcommand{\be}{\begin{equation}}
\newcommand{\ee}{\end{equation}}

\title{Modeling the black holes surrounded by a dark matter halo in the galactic center of M87}

\author[a]{Dong Liu,}
\author[b]{Yi Yang,}
\author[a]{Zhaoyi Xu}
\author[a,1]{and Zheng-Wen Long\note{Corresponding author.}}

\affiliation[a]{College of Physics, Guizhou University, Guiyang, 550025, China}
\affiliation[b]{School of Mathematics and Statistics, Guizhou University of Finance and Economics, Guiyang, 550025, China}

\emailAdd{dongliuvv@yeah.net}
\emailAdd{yangyigz@yeah.net}
\emailAdd{zyxu@gzu.edu.cn}
\emailAdd{zwlong@gzu.edu.cn}

\abstract{In this paper, the structure of a dark matter halo can be well described by the mass model of M87 and the Einasto profile for the cold dark matter model, i.e., $\rho_{\text{eina}} (r)=\rho_\text{e} \exp ( -2 \alpha ^{-1} ((r/r_\text{e})^\alpha -1 ) )$ \href{https://doi.org/10.1038/s41586-020-2642-9}{(Wang et al. in Nature \textbf{585}:39-42, 2020)}. Under these conditions, we construct a solution of a static spherically symmetric black hole in a dark matter halo. Then, using the Newman-janis algorithm, we extend this static solution to the case of rotation, and obtain a solution for the Kerr-like black hole. We prove that this solution of the Kerr-like black hole is indeed a solution to the Einstein field equations. Finally, taking M87 as an example, we study and analyze some physical properties of this Kerr-like black hole, and then compare them with the Kerr black hole. Particularly, from the perspective of the black hole shadow and the fact that the Kerr-like black hole and the Kerr black hole is distinguishable, we give the upper limit of the shape parameter of the Einasto density profile, that is approximately $\alpha<0.22$, which may provide a new method to further improve and perfect the density profile of dark matter model. These research results for the black hole in a dark matter halo may indirectly provide an effective method for detecting the existence of dark matter.}


\begin{document}
\maketitle
\flushbottom

\section{Introduction}\label{s1}
Black hole is a fascinating and mysterious celestial body predicted by General Relativity. General Relativity predicts that a sufficiently compact mass can warp spacetime, forming the gravitational field. In other words, when the mass of the celestial body is large enough, the celestial body will make the gravitational field so strong that it can prevent light from escaping, and eventually a black hole may be emerged. The famous physicists K. Schwarzschild and R. Kerr respectively solved the Einstein field equations of the vacuum, and then  they obtained the well-known the exact solution of the Schwarzschild black hole \cite{Schwarzschild:1916uq} and Kerr black hole \cite{Kerr:1963ud}. In 2019, the Event Horizon Telescope (EHT) released the first image of a black hole, that is a supermassive black hole at the center of M87 in the Virgo elliptical galaxy \cite{EventHorizonTelescope:2019dse}. The photo of the black hole is composed of two parts, that is the outer circular bright area and the inner dark area. The bright region near the black hole represents the accretion disk, which appears distorted due to gravitational lensing. And the side of the detector away from the accretion disk will be relatively darker, forming a dark region. The images of the black hole shadow can help us to understand the spacetime geometry of the event horizon for the black hole and its speed of rotation. From these Refs. \cite{Okyay:2021nnh,Churilova:2021tgn,Ovgun:2018tua,Kumar:2019pjp,Bambi:2019tjh,Konoplya:2021slg,Khan:2019gco,He:2020dfo,Kuang:2022xjp,Kuang:2022ojj,Perlick:2021aok,Wei:2020ght,Roy:2021uye,Chen:2022nbb,Tang:2022uwi,Yang:2023agi,Capozziello:2023tbo}, we can learn more about the study of the black hole shadow. In 2022, the EHT Collaboration released the first photo of Sgr A, that is the supermassive black hole in the center of the Milky Way \cite{EventHorizonTelescope:2022apq}. Therefore, it is generally accepted for us that black hole exists in our universe.

On the other hand, more and more observational data indicate the existence of dark matter, such as the rotation curves of spiral galaxies, the cosmic microwave background radiation and gravitational lensing, etc. Based on these observational data, astronomers have proposed a series of dark matter models to study dark matter, such as cold dark matter (CDM) model \cite{Navarro:1995iw,Navarro:1996gj}, warm dark matter (WDM) model \cite{Dodelson:1993je}, scalar field dark matter (SFDM) model \cite{Urena-Lopez:2002ptf} and self-interacting dark matter (SIDM) model \cite{Spergel:1999mh}. In these dark matter models, the density profile of dark matter can always be well described by relatively few variable parameters, such as the density parameter $\rho$ and the characteristic radius $r$. So far, various density profiles of dark matter models have been proposed, and we list some of them as follows. F. Navarro et al. gave the NFW profile for studying the dark matter halo structure in the CDM model \cite{Navarro:1995iw}. Then, T. Matos et al. gave the spacetime geometry of the pure dark matter halo for this NFW profile \cite{Matos:2003nb}. L. Urena-Lopez et al. gave the density profile of scalar field dark matter model by solving Einstein Klein-Gordon equation coupled with the scalar field \cite{Urena-Lopez:2002ptf}. A. Burkert gave the density of the URC profile from the dark matter halos of seven dwarf spiral galaxies \cite{Burkert:1995yz}. The accuracy of the density profile of dark matter may directly determine the difficulty of finding dark matter. Here, we main focus on the density profile proposed by J. Einasto for the CDM model \cite{1965On}. J. Wang et al. showed in their recent research that the NFW profile and the Einasto profile can well describe the structure of dark matter halo in the entire range of the mass \cite{Wang:2019ftp}.  More density profiles and models of dark matter can be found in Refs. \cite{Salucci:2018hqu,Matos:2023usa}. These studies do make positive contributions to the indirect search for dark matter.

Based on these interesting physical backgrounds above, a large number of the individuals are beginning to study the interaction between black holes and dark matter. If there is dark matter near the black hole, due to the strong gravity of the black hole itself, the dark matter will be distributed in a certain structure near the black hole and disappear at a certain distance away from the black hole. Using Newtonian approximations, P. Gondolo et al. first obtained the density profile of dark matter near black hole at the center of the Milky Way \cite{Gondolo:1999ef}. L. Sadeghian et al. used accurate Schwarzschild geometry to perform relativistic calculations, and corrected the density profile of P. Gondolo's dark matter \cite{Sadeghian:2013laa}. Xu et al. combined the NFW profile in the dark matter halo structure and the spike structure, and then they obtained the spacetime geometry of the black hole immersed in dark matter \cite{Xu_2018,Xu:2021dkv,Xu:2020jpv}. S. Nampalliwar et al. tested the impacts of dark matter spike on the black hole at the center of the Milky Way \cite{Nampalliwar:2021tyz}. Ning Dai et al. analyzed the possibility of the existence of  a Schwarzschild black hole immersed in a dark matter halo with Hernquist type density distribution from the perspective of gravitational waves \cite{Dai:2023cft}. Combining the previously mentioned photos of the black holes at the center of the M87 and Sgr A, the authors began to study these black holes in a dark matter halo of M87 and Sgr A and we list some below. K. Jusuf et al. obtained the spacetime geometry of the URC profile dark matter in the center of the M87 galaxy, and studied the impacts of dark matter parameters on the shadow of the black hole \cite{Jusufi:2019nrn}. H. Davoudiasl et al. used the Event Horizon Telescope to measure the ultralight bosons in M87 and then study the distribution of dark matter on the large-scale structure \cite{Davoudiasl:2019nlo}. A. K. Saha et al. used the Event Horizon Telescope to detect the ultralight axions of Sgr A, and thus gave the mass limit of ultralight bosons \cite{Saha:2022hcd}. V. Luca et al. calculated the density profile of superfluid dark matter near the supermassive black hole at the center of the galaxy \cite{DeLuca:2023laa}. Z. Shen et al. proposed five analytical models of dark matter halo near the supermassive black hole, and gave the corresponding black hole solutions in dark matter halo \cite{Shen:2023erj}. N. Molla et al. used strong gravitational lensing to study black holes in dark matter halo of M87 and Sgr A \cite{Molla:2023yxn}. V. Xavier et al. used black hole shadows to study black holes in dark matter halo of M87 \cite{Xavier:2023exm}. At the same time, a recent article published in \emph{Nature} once again showed that the NFW profile and the Einasto profile of dark matter can well describe the astronomically predicted the structure of dark matter halo \cite{Wang:2019ftp}. For the studies of the Einasto profiles and black holes, we list some below. V. Cardoso et al. studied and explored black holes immersed in the Einasto profile from geodesic equations and axial gravitational perturbations \cite{Figueiredo:2023gas}. F. Rahaman et al. obtained a solution of a transmissible wormhole in the Einasto profile from Einstein's field equations \cite{Rahaman:2023tkm}. If it is said that there is dark matter near the black hole, then this dark matter must be able to change the spacetime geometry of the black hole. Therefore, it is meaningful to study the impacts of dark matter halo on the spacetime geometry of the black hole. Therefore, in this work, we will study and explore the impacts of the Einasto profile of dark matter on the black hole in the center of M87. In addition, over the years, the gravitational waves produced by the merger of binary black holes have also been gradually detected by LIGO/Virgo \cite{LIGO1,LIGO2,LIGO3,LIGO4}. The source of gravitational waves is usually due to perturbations in the black hole or additional sources such as dark matter. Different gravitational waves may be generated according to different black holes, and we can use this to judge the type of black hole or the background space-time. V. Vishveshwara et al. indicated that a black hole in the perturbed state can emit gravitational waves \cite{Vishveshwara:1970zz}. And this gravitational wave is usually dominated by the excited oscillation mode of complex frequency, that is quasinormal mode (QNM) \cite{Moderski:2005hf}.  As a characteristic sound, the QNM from the black hole can provide us a new method to identify the black hole in our universe. The recent studies about quasinormal modes of the black hole in the dark matter can refer to these Refs. \cite{Zhang:2021bdr,Zhang:2022roh,Liu:2021xfb,Liu:2022ygf,Konoplya:2022hbl}. In the future, with the increase of LIGO/Virgo observation data and the photos of the black hole shadow, it is possible to detect such black holes in a dark matter halo, which also provides an effective method for indirect detection of the existence of dark matter. Furthermore, from the geometry of the black hole shadow, we can define observables and establish a connection between them and black hole parameters \cite{Hioki:2009na,Lee:2021sws,Ghosh:2020ece,Tang:2022uwi}. In other words, if we observe a shadow image of a black hole in our universe, we can deduce the relevant parameters of the black hole according to the model, even the metric of the black hole. S. Vagnozzi et al. studied the impact of the observables of black hole shadows on some general black holes in M87 and Sgr A, which has received widespread attention \cite{Vagnozzi:2019apd,Allahyari:2019jqz,Vagnozzi:2022moj,Afrin:2022ztr,Khodadi:2020jij}. They show that the observables of black hole shadows can, to a certain extent, be used to test the black hole parameters or the parameters of physical models. These results may be expected to promote the development of physics. On the other hand, for some black holes coupled with other matter, such as dark matter or dark energy \cite{Contreras:2021yxe}, these observables of the black hole shadow may help to examine the density profile in a dark matter halo, or the models of dark matter and dark energy. In this work, we are considering black holes coupled with dark matter, and we will try to study these dark matter profiles from the perspective of black hole shadows, which may provide a new method for verifying and improving the density profile of dark matter and even the dark matter model. This is also one of the research objectives of this work.

This paper is organized as follows. In section \ref{s2}, we introduce a much-concerned Einasto profile that can well describe the cold dark matter model, and then we give the metric for this pure dark matter spacetime. In section \ref{s3}, starting from Einstein field equations, we introduce a spherically symmetric black hole into this pure dark matter spacetime, and then we give the metric of the Schwarzschild-like black hole in a dark matter halo. In section \ref{s4}, using the Newman-Janis algorithm, we extend the solution of this Schwarzschild-like black hole to the case of the Kerr-like black hole. Besides, in Appendix \ref{sa}, we prove that this solution of the Kerr-like black hole is a solution of Einstein field equations. In section \ref{s5}, we study and analyze some basic physical properties of this Kerr-like black hole in a dark matter halo, and then we compare these results with the Kerr black hole. Here, we list the following physical properties: 1) The horizons of the Kerr-like black hole; 2) The ergosphere of the Kerr-like black hole; 3) The motion equation of neutral particle near the Kerr-like black hole; 4) The geodesic equations of lightlike particle and photon regions; 5) The shadow of the Kerr-like black hole in a dark matter halo. 6) The shadow observables of the Kerr-like black hole in a dark matter halo. Finally, section \ref{s6} is our conclusion and discussion. The relevant parameters of the galaxy M87 are $\rho_\text{e}=6.9 \times 10^6 M_\odot/kpc^3, r_\text{e}=91.2kpc$ and the mass of the black hole at the center of the M87 is $6.5 \times 10^9 M_\odot$ \cite{EventHorizonTelescope:2019dse}. In this paper, we mainly use the black hole units of $G=c=M_\text{SBH}=1$, the radius of the Schwarzschild black hole (SBH) is given by $r_{\text{SBH }}=2 GM_\text{SBH}/c^2$.

\section{Density profile of dark matter and its spacetime geometry}\label{s2}
Recently, one density profile firstly proposed by J. Einasto has received widespread attention \cite{1965On}. More studies on this density profile of dark matter can be found in these Refs. \cite{Marzola:2016hyt,Brdar:2016ifs,Iorio:2013ida,Huang:2016pxg,DAMPE:2021hsz}. Particularly, J. Wang et al. use this density profile to fit the observed data of the halo masses (over 20 orders of magnitude), and they find that the observed data can be well described by this density profile of the simple two-parameter \cite{Wang:2019ftp}. This density profile is also known as the Einasto profile, and it reads
\begin{equation}
\rho_{\text{eina}} (r)=\rho_\text{e} \exp [ -2 \alpha ^{-1} ((\frac{r}{r_\text{e}})^\alpha -1 ) ],
\label{e21}
\end{equation}
where, $r_\text{e}$ is the characteristic radius and its logarithmic slope is $\text{d ln}  \rho / \text{d ln} r=-2$, $\rho_\text{e}$ is the halo density and $\alpha$ is a shape parameter\footnote{The authors in Ref. \cite{Wang:2019ftp} fixed this shape parameter $\alpha$ at $\alpha=0.16$, but others are not \cite{Udrescu:2018hvl,Quintana:2022yky,Figueiredo:2023gas,Rahaman:2023tkm}. Therefore, in our work, we reserve $\alpha$ as a free parameter to participate in the calculation later.}. Based on the density profile (\ref{e21}), the mass distribution of the dark matter halo is given by 
\begin{equation}
\begin{aligned}
M_{\text{eina}}(r) = 4\pi \int_{0}^{r}\rho_{\text{eina}} (r')r'^2dr' =\frac{4\pi e^{2/\alpha } \rho_\text{e} \left ( 8^{-1/\alpha } r_\text{e}^3 \alpha ^{3/\alpha }\Gamma [\frac{3}{\alpha }] -r^3 \text{Ei}[\frac{-3 + \alpha}{\alpha },\frac{2}{\alpha }(\frac{r}{r_\text{e}})^\alpha] \right ) }{\alpha },
\label{e22}
\end{aligned}
\end{equation}
where, $\text{Ei}[n,z]$ is the exponential integral function $\text{Ei}_n(z)$ and $\Gamma[z]$ is the gamma function $\Gamma(z)$. According to Newtonian theory, for a test particle in the equatorial plane of spherical symmetric spacetime, its tangential velocity can be determined by the mass distribution of dark matter \cite{Boehmer:2007um}. Therefore, the tangential velocity $V$ can be defined as
\begin{equation}
\begin{aligned}
V_{\text{eina}}(r) = \sqrt{\frac{M_\text{eina}(r)}{r}}= \sqrt{4\pi e^{2/\alpha } \rho_\text{e} \left ( 8^{-1/\alpha } r_\text{e}^3 \alpha ^{3/\alpha }\Gamma [\frac{3}{\alpha }]/r -r^2 \text{Ei}[\frac{-3 + \alpha}{\alpha },\frac{2}{\alpha }(\frac{r}{r_\text{e}})^\alpha] \right )/\alpha  }.
\label{e23}
\end{aligned}
\end{equation}
On the other hand, for the pure dark matter spacetime, its metric can always be written as a static spherical symmetric form
\begin{equation}
\begin{aligned}
ds^{2}=-f(r)dt^{2}+\frac{1}{g(r)}dr^{2}+r^{2}(d\theta ^{2}+\sin^{2}\theta d\phi ^{2}),
\label{e24}
\end{aligned}
\end{equation}
where, $f(r)$ is the redshift function and $g(r)$ is the shape function. Here, we mainly consider the simpler case in the spherical symmetric spacetime, that is, $f(r)=g(r)$. The relationship between the redshift function $f(r)$ and the tangential velocity $V$ suggested in Ref. \cite{Matos:2003nb}:
\begin{equation}
V_{\text{eina}}(r)^2 = \frac{r}{\sqrt{f(r)}}\frac{d\sqrt{f(r)}}{dr}=r\frac{dln\sqrt{f(r)}}{dr}.
\label{e25}
\end{equation}
According to the definitions  (\ref{e23}) and (\ref{e25}), we can get an analytical solution of the metric coefficient $f(r)$ in a dark matter halo, and it finds that
\begin{equation}
f_{\text{eina}}(r)=\exp\left (- \frac{4\pi 2^{1-3/\alpha } e^{2/\alpha } r^2 ((r/r_\text{e})^{\alpha }/\alpha )^{-3/\alpha }\rho_\text{e} \Gamma [\frac{3}{\alpha },0,\frac{2(\frac{r}{r_\text{e}})^\alpha} {\alpha} ]}{\alpha}  \right ),
\label{e26}
\end{equation}
where, $\Gamma[a,z_0,z_1]$ is the generalized incomplete gamma function as $ \Gamma[a,z_0,z_1] = \Gamma(a,z_0)-\Gamma(a,z_1)$, and then the incomplete gamma function $\Gamma [a,z]$ can be defined by an integral expression as $\Gamma [a,z]=\int _{z}^{\infty } t^{a-1}e^{-t}dt $. As an example, we consider the galaxy of M87 and its relevant parameter are $\rho_\text{e}=6.9 \times 10^6 M_\odot/kpc^3, r_\text{e}=91.2kpc$. The mass of the black hole in the center of M87 is $M_{\text{BH}} = 6.5 \times 10^{9}M_{\bigodot}$.  In this case, the dark matter parameters can be converted in the black hole units (BHU) by the following formula: $r_\text{e}(\text{BHU})=r_\text{e} / (2 G M_{\text{BH}}/c^2) \times r_\text{BH}$ and $\rho_\text{e}(\text{BHU})=\rho_\text{e}/(M_\text{BH}/(4/3 \pi (2 G M_{\text{BH}}/c^2)^3))\times \rho_{\text{BH}}$. For the $f(r)$ in function (\ref{e26}), it is not difficult to find that
\begin{equation}
\lim_{\rho_e\rightarrow 0}f_{\text{eina}}(r)=1,\lim_{r\rightarrow \infty }f_{\text{eina}}(r)=1,
\label{e27}
\end{equation}
The Eqs. (\ref{e27}) show that when the density of dark matter is $0$ or the distance is infinity away from the black hole, the dark matter is absent. Therefore, the spacetime return to Minkowski flat spacetime.

\section{Schwarzschild-like black hole in a dark matter halo}\label{s3}
In this section, we extend our results (\ref{e26}) to the case of black hole in a dark matter halo. We will strictly follow the method recorded in the Refs. \cite{Xu_2018,Xu:2021dkv}. First of all, we need to solve the Einstein field equation of pure dark matter spacetime, which has the following form
\begin{equation}
R_{\mu \nu }-\frac{1}{2}g_{\mu \nu }R=\kappa ^2 T_{\mu \nu }(\text{DM-halo}),
\label{e31}
\end{equation}
where, $R_{\mu \nu }$ is Ricci tensor, $g_{\mu \nu }$ is the metric of pure dark matter spacetime and $R$ is Ricci scalar. The energy momentum tensor of this spacetime can be defined as ${T^\nu}_\mu=g^{\nu \alpha }T_{\mu \alpha }=diag[-\rho ,p_r, p, p]$, and $g^{\nu \alpha }$ is the inverse metric of the spacetime. Taking the metric (\ref{e24}) into Eq. (\ref{e31}), we can obtain the following equations
\begin{equation}
\begin{aligned}
&\kappa ^2 {T^{t}}_{t}(\text{DM-halo})=g(r)\left ( \frac{1}{r}\frac{g'(r)}{g(r)}+\frac{1}{r^2} \right ) - \frac{1}{r^2}, \\
&\kappa ^2 {T^{r}}_{r}(\text{DM-halo})=g(r)\left ( \frac{1}{r^2} + \frac{1}{r}\frac{f'(r)}{f(r)} \right )-\frac{1}{r^2},\\
&\kappa ^2 {T^{\theta }}_{\theta }(\text{DM-halo})=\kappa ^2 {T^{\phi }}_{\phi }(\text{DM-halo})\\
&=\frac{1}{2}g(r)\left ( \frac{f''(r)f(r)-f'(r)^2}{f(r)^2}+\frac{f'(r)^2}{2f(r)^2} +\frac{1}{r}(\frac{f'(r)}{f(r)}+\frac{g'(r)}{g(r)}) + \frac{f'(r)g'(r)}{2f(r)g(r)}\right ).
\label{e32}
\end{aligned}
\end{equation}
When the black hole is in a dark matter halo, then the energy momentum tensor can be written as ${T^\nu}_\mu={T^\nu}_\mu(\text{BH}) +{T^\nu}_\mu(\text{DM-halo})$. For Schwarzschild black hole, the energy momentum tensor is $0$, that is, ${T^\nu}_\mu(\text{BH})=0$. That is to say, we only need to consider the energy-momentum tensor ${T^\nu}_\mu(\text{DM-halo})$ of the dark matter halo. Therefore, based on the above analysis, we can assume the new metric of black hole in a dark matter halo that
\begin{equation}
\begin{aligned}
ds^{2}=-(f(r)+F1(r))dt^{2}+\frac{1}{g(r)+G1(r)}dr^{2}+r^{2}(d\theta ^{2}+\sin^{2}\theta d\phi ^{2}),
\label{e33}
\end{aligned}
\end{equation}
where, $f(r)$ and $g(r)$ are the coefficients of the pure dark matter halo. Then Eq. (\ref{e31}) can be written as
\begin{equation}
R_{\mu \nu }-\frac{1}{2}g_{\mu \nu }R=\kappa ^2 (T_{\mu \nu }(\text{BH})+T_{\mu \nu }(\text{DM-halo})).
\label{e34}
\end{equation}
Taking the metric (\ref{e33}) into Eq. (\ref{e34}) and then comparing with Eq. (\ref{e32}), we can obtain
\begin{equation}
\begin{aligned}
&(g(r)+G1(r))\left ( \frac{1}{r^2}+\frac{1}{r}\frac{g'(r)+G1'(r)}{g(r)+G1(r)} \right ) =g(r)\left ( \frac{1}{r^2}+\frac{1}{r}\frac{g'(r)}{g(r)} \right ),\\
&(g(r)+G1(r))\left ( \frac{1}{r^2} + \frac{1}{r}\frac{f'(r)+F1'(r)}{f(r)+F1(r)} \right )=g(r)\left ( \frac{1}{r^2} + \frac{1}{r}\frac{f'(r)}{f(r)}\right ).
\label{e35}
\end{aligned}
\end{equation}
Eq. (\ref{e35}) is the energy-momentum tensors that can be described by two different metrics. The first equation of Eq. (\ref{e35}) is only related to function $G1$. The second equation is related to functions $G1,F1$ and it can be simplified as
\begin{equation}
\begin{aligned}
\frac{f'(r)+F1'(r)}{f(r)+F1(r)} = \frac{g(r)}{g(r)+G1(r)}\left ( \frac{1}{r} + \frac{f'(r)}{f(r)}\right ) -\frac{1}{r}.
\label{e36}
\end{aligned}
\end{equation}
The general solution to the first equation of (\ref{e35}) is $G1(r)=C/r$. $C$ is an undetermined coefficient, which needs to be determined by boundary condition. Inspired by the Eq. (\ref{e27}), here, we use the Schwarzschild black hole as the boundary condition. Under this boundary condition, the new metric $G(r)=g(r)+G1(r)$ of the black hole in a dark matter halo will degenerate into the Schwarzschild black hole at the boundary ($G(r)=1-2M /r$). At this point, a particular solution that satisfies the boundary conditions is g(r)=1, and then $G1(r)=-2M/r$. In the end, we determined that the constant C is $C=-2M$. Therefore, we can obtain the analytical expressions of functions $F1, G1$, which are as follows
\begin{equation}
\begin{aligned}
&F1(r)=\exp\left ( \int \frac{g(r)}{g(r)-2M_{\text{BH}}/r}\left ( \frac{1}{r}+\frac{f'(r)}{f(r)} \right )dr-\frac{1}{r}dr \right )-f(r),\\
&G1(r)=-\frac{2M_{\text{BH}}}{r},
\label{e37}
\end{aligned}
\end{equation}
where, $M_\text{BH}$ is the mass of the black hole. Here, using the case of $f(r)=g(r)$ we assumed before in Eq. (\ref{e35}), it is not difficult to find that $F1(r)=G1(r)=-2M_\text{BH}/r$. Therefore, the new metric for the Schwarzschild-like black hole in a dark matter halo is given by
\begin{equation}
\begin{aligned}
ds^{2}=-F(r)dt^{2}+\frac{1}{G(r)}dr^{2}+H(r)(d\theta ^{2}+\sin^{2}\theta d\phi ^{2}),
\label{e38}
\end{aligned}
\end{equation}
where, $F(r)=f(r)+F1(r), G(r)=g(r)+G1(r), H(r)=r^{2}$. The terms $f(r)$ and $g(r)$ represent the factor terms for considering pure dark matter halo in the metric (\ref{e26}). Finally, the metric coefficients for the Schwarzschild-like black hole in a dark matter halo are as follows
\begin{equation}
F(r)=G(r)=\exp\left (- \frac{4\pi 2^{1-3/\alpha } e^{2/\alpha } r^2 ((r/r_\text{e})^{\alpha }/\alpha )^{-3/\alpha }\rho_\text{e} \Gamma [\frac{3}{\alpha },0,\frac{2(\frac{r}{r_\text{e}})^\alpha} {\alpha} ]}{\alpha}  \right )-\frac{2M_{\text{BH}}}{r}.
\label{e39}
\end{equation}
If the dark matter is absent, that is, $\rho_\text{e}=0$, and then the metric (\ref{e39}) degenerates into the case of the Schwarzschild black hole.

\section{Kerr-like black hole in a dark matter halo}\label{s4}
In the previous section, we derived the metric of Schwarzschild-like black hole in a dark matter halo from the Einasto profile. Next, we will extend this Schwarzschild-like black hole metric to the Kerr-like metric. The method we used is Newman-Janis (N-J) algorithm, which is one of the commonly used and efficient methods \cite{Newman:1965tw,Azreg-Ainou:2014pra,Jusufi:2019nrn,Kim:2021vlk,Xu:2021dkv}. Strictly following this method, we firstly transform the metric (\ref{e38}) from Boyer-Lindquist (BL) coordinates $(t, r,\theta,\phi)$ to Eddington-Finkelstein (EF) coordinates ($u, r, \theta, \phi$)
\begin{equation}
du=dt -\frac{dr}{\sqrt{F(r)G(r)}},
\label{e41}
\end{equation}
and then one can obtain
\begin{equation}
ds^2=-F(r)du^2 -2\sqrt{\frac{F(r)}{G(r)}}du dr+H(r)(d\theta ^2 + \sin^2\theta d\phi ^2),
\label{e42}
\end{equation}
where, $F(r), G(r), H(r)=r^2$ are the metric functions. In the null tetrad, the inverse metric of Schwarzschild-like black hole can be composed of basis vectors $l^{\mu},n^{\mu},m^{\mu},\bar{m}^{\mu}$, and it is given by
\begin{equation}
g^{\mu\nu}=-l^{\mu}n^{\nu}-l^{\nu}n^{\mu}+m^{\mu}\bar{m}^{\nu}+m^{\nu}\bar{m}^{\mu},
\label{e43}
\end{equation}
where, these basis vectors $l^{\mu},n^{\mu},m^{\mu},\bar{m}^{\mu}$ are defined as follows
\begin{equation}
\begin{aligned}
&l^{\mu}=\delta _r^\mu,\\
&n^\mu=\sqrt{\frac{G(r)}{F(r)}}\delta _u^\mu-\frac{1}{2}G(r)\delta _r^\mu,\\
&m^\mu=\frac{1}{\sqrt{2}\times r}\left ( \delta _\theta ^\mu + \frac{i}{\sin{\theta }}\delta _\phi ^\mu \right ),\\
&\bar{m}^\mu=\frac{1}{\sqrt{2}\times r}\left ( \delta _\theta ^\mu - \frac{i}{\sin{\theta }}\delta _\phi ^\mu \right ),
\label{e44}
\end{aligned}
\end{equation}
where, the basic vectors $l$ and $n$ are real and basic vectors $m$ and $\bar{m}$ are mutually complex conjugate \cite{CiriloLombardo:2004qw}. The relationship between these basis vectors satisfies the conditions of normalization, orthogonality and isotropy. According to the Newman-Janis algorithm, the spacetime coordinates between different observers satisfy the complex transformation. Therefore, we can write the new coordinates as follows
\begin{equation}
\begin{aligned}
x'^\mu=x^\mu - ia(\delta _u^\mu - \delta^\mu_r )\cos\theta \rightarrow \left\{\begin{matrix}\begin{aligned}
&u'=u-ia\cos\theta \\
&r'=r+ia\cos\theta \\
&\theta '=\theta \\
&\phi '=\phi
\end{aligned}
\end{matrix}\right.
\label{e45}
\end{aligned}
\end{equation}
where, $a$ is rotation parameter, $x'^\mu$ and $x^\mu$ are the EF coordinate components of the case of rotation and non-rotation, respectively. Under this condition, these new metric functions can be rewritten as $F(r)\rightarrow\mathcal{F}(r, a,\theta)$, $G(r)\rightarrow\mathcal{G}(r, a, \theta)$, $H(r)\rightarrow \Sigma(r, a, \theta)$. Besides, in the case of rotation, a new null tetrad $Z'^\mu$ can be transformed from the null tetrad of non-rotation through the relation $Z'^\mu=(\partial x'^\mu /\partial x^\mu)Z^\mu$ \cite{Drake:1998gf,Jusufi:2019nrn}. Therefore, we can obtain this rotation null tetrad with the new metric functions $\mathcal{F}(r, a, \theta)$, $\mathcal{G}(r, a, \theta)$ and $\Sigma(r, a, \theta)$ similar to the definitions (\ref{e44}),
\begin{equation}
\begin{aligned}
l'^{\mu}& = \frac{\partial x'^\mu}{\partial x^\mu}l^\mu=1\times\delta _r^\mu=\delta _r^\mu,\\
n'^\mu& = \frac{\partial x'^\mu}{\partial x^\mu}n^\mu=1 \times \left(\sqrt{\frac{\mathcal{G}}{\mathcal{F}}}\delta _u^\mu-\frac{1}{2}\mathcal{G}\delta _r^\mu\right)=\sqrt{\frac{\mathcal{G}}{\mathcal{F}}}\delta _u^\mu-\frac{1}{2}\mathcal{G}\delta _r^\mu,\\
m'^\mu& = \frac{\partial x'^\mu}{\partial x^\mu}m^\mu=\left ( 1-ia(\delta _u^\mu-\delta _r^\mu)\frac{\partial \cos \theta}{\partial \theta} \right )\frac{1}{\sqrt{2}\Sigma_1}\left ( \delta _\theta ^\mu+\frac{i}{\sin \theta}\delta _\phi ^\mu \right )\\
&=\frac{1}{\sqrt{2 }\Sigma_ 1}\left ( \delta _\theta ^\mu+\frac{i}{\sin \theta}\delta _\phi ^\mu -ia(\delta _u^\mu-\delta _r^\mu)\frac{\partial \cos \theta }{\partial \theta} \times \left ( \delta _\theta ^\theta+\frac{i}{\sin \theta}\delta _\phi ^\theta \right ) \right )\\
&=\frac{1}{\sqrt{2}\Sigma_ 1}\left ( \delta _\theta ^\mu +ia\sin\theta (\delta _u^\mu-\delta _r^\mu)+ \frac{i}{\sin{\theta }}\delta _\phi ^\mu \right ),\\
\bar{m}'^\mu& = \overline{m'^\mu}=\frac{1}{\sqrt{2 }\Sigma_2}\left ( \delta _\theta ^\mu - ia\sin\theta (\delta _u^\mu-\delta _r^\mu)- \frac{i}{\sin{\theta }}\delta _\phi ^\mu \right ),
\label{e46}
\end{aligned}
\end{equation}
where, $\overline{m'^\mu}$ means the complex conjugate of $m'^\mu$, $\Sigma_1 =r-ia\cos \theta$ and $\Sigma_2=r + ia\cos \theta$. Taking the results of (\ref{e46}) into the equation of new contravariant metric,
\begin{equation}
\begin{aligned}
g'^{\mu\nu}=-l'^{\mu}n'^{\nu}-l'^{\nu}n'^{\mu}+m'^{\mu}\bar{m}'^{\nu}+m'^{\nu}\bar{m}'^{\mu}, 
\end{aligned}
\end{equation}
and then we can obtain the contravariant non-zero components of this new metric which are as follows,
\begin{equation}
\begin{aligned}
&g'^{u u}=\frac{a^2\sin^2\theta}{\Sigma },g'^{r r}=\mathcal{G}+\frac{a^2\sin^2\theta}{\Sigma },g'^{\theta \theta }=\frac{1}{\Sigma },g'^{\phi \phi }=\frac{1}{\Sigma \sin^2\theta },\\
&g'^{ur}=g'^{ru}=-\sqrt{\frac{\mathcal{G}}{\mathcal{F}}}-\frac{a^2\sin ^2 \theta}{\Sigma },g'^{u\phi }=g'^{\phi u}=\frac{a}{\Sigma },g'^{r\phi }=g'^{\phi r}=-\frac{a}{\Sigma},
\label{e47}
\end{aligned}
\end{equation}
where, $\Sigma=\Sigma_1\Sigma_2=r^2+a^2\cos^2\theta$. Based on these inverse metric and then taking the inverse of (\ref{e47}), the new metric of the Kerr-like black hole in EF coordinates becomes
\begin{equation}
\begin{aligned}
ds^{2}=&-\mathcal{F}du^{2}-2\sqrt{\frac{\mathcal{F}}{\mathcal{G}}}dudr+2a\sin^2\theta \left (\mathcal{F}- \sqrt{\frac{\mathcal{F}}{\mathcal{G}}} \right )dud\phi +2a\sin^2\theta \sqrt{\frac{\mathcal{F}}{\mathcal{G}}}drd\phi \\
&+\Sigma d\theta ^2  + \sin^2\theta \left [ \Sigma +a^2\sin^2\theta \left ( 2\sqrt{\frac{\mathcal{F}}{\mathcal{G}}} - \mathcal{F} \right ) \right ]d\phi ^2.
\label{e48}
\end{aligned}
\end{equation}
Finally, we only need to convert this EF coordinates into the coordinates we are familiar with, that is, BL coordinates. Here, we follow the method from the Ref. \cite{Azreg-Ainou:2014pra}, which introduces more physical parameters and symmetry properties, simplifying the complex procedure of the Newman-Janis algorithm. This method is further studied and applied in these Refs. \cite{Toshmatov:2015npp,Jusufi:2019nrn,Wei:2020ght,Nampalliwar:2021tyz,Kumar:2020owy}. Firstly, we introduce the following transformations,
\begin{equation}
\begin{aligned}
du=dt-\frac{K(r)+a^2}{G(r)H(r)+a^2 }dr, d\phi =d\varphi -\frac{a}{G(r)H(r)+a^2 }dr,
\label{e49}
\end{aligned}
\end{equation}
where, $K(r)=H(r)\sqrt{G(r)/F(r)}$ and
\begin{equation}
\begin{aligned}
\mathcal{F}(r,a,\theta)=\frac{(G(r)H(r)+a^2\cos^2\theta)\Sigma }{(K(r)+a^2\cos^2\theta)^2},
\mathcal{G}(r,a,\theta)=\frac{G(r)H(r)+a^2\cos^2\theta }{\Sigma },
\label{e410}
\end{aligned}
\end{equation}
and then taking (\ref{e49}), (\ref{e410}) into (\ref{e48}), we can obtain the final form of metric under BLC coordinates
\begin{equation}
\begin{aligned}
ds^2= & -\frac{\left(G H+a^2 \cos ^2 \theta\right) \Sigma}{\left(K+a^2 \cos ^2 \theta\right)^2} dt^2+\frac{\Sigma}{G H+a^2} dr^2-2 a \sin ^2 \theta\left[\frac{K-G H}{\left(K+a^2 \cos ^2 \theta\right)^2}\right] \Sigma dt d \phi \\
&+ \Sigma d\theta^2  +\Sigma \sin ^2 \theta\left[1+a^2 \sin ^2 \theta \frac{2 K-G H+a^2 \cos ^2 \theta}{\left(K+a^2 \cos ^2 \theta\right)^2}\right] d\phi^2,
\label{e411}
\end{aligned}
\end{equation}
where, $G, H, K$ are the function of $r$ we introduced before and $\Sigma=r^2+a^2\cos^2\theta$. With the case $F(r)=G(r), H(r)=r^2$ we introduced previous section, we find that $K(r)=H(r)=r^2$. Finally, the metric (\ref{e411}) can be written as Kerr-like form,
\begin{equation}
\begin{aligned}
ds^2= & -(1-\frac{r^2-g(r)r^2+2Mr}{\Sigma }) dt^2 +\frac{\Sigma}{\Delta } dr^2 -2 a \sin ^2 \theta(\frac{r^2-g(r)r^2+2Mr}{\Sigma }) dt d \phi \\
&+ \Sigma d\theta^2 +\left ( (a^2+r^2)\sin^2 \theta +\frac{a^2\sin^4\theta(r^2-g(r)r^2+2Mr)}{\Sigma }\right ) d\phi^2,
\label{e412}
\end{aligned}
\end{equation}
where,
\begin{equation}
\begin{aligned}
\Delta& = r^2G(r)+a^2=r^2(g(r)-2M/r)+a^2\\
&=\exp\left ( \frac{2\alpha\text{ln}r-4\pi 2^{1-3/\alpha } e^{2/\alpha } r^2 ((r/r_\text{e})^{\alpha }/\alpha )^{-3/\alpha }\rho_\text{e} \Gamma [\frac{3}{\alpha },0,\frac{2(\frac{r}{r_\text{e}})^\alpha} {\alpha} ]}{\alpha}  \right )-2Mr+a^2.
\label{e413}
\end{aligned}
\end{equation}
It is easy to find that in (\ref{e412}), if the dark matter is absent, (i.e., $\rho_\text{e}=0$, and then $g(r)=1$)), the metric of black hole in a dark matter halo degenerates into the Kerr metric
\begin{equation}
\begin{aligned}
ds^2=&  -(1-\frac{2Mr}{\Sigma }) dt^2 +\frac{\Sigma}{\Delta } dr^2 -\frac{4Mr a  \sin ^2 \theta }{\Sigma } dt d \phi + \Sigma d\theta^2 \\
&+\left ( (a^2+r^2)\sin^2 \theta +\frac{2Mr a^2\sin^4\theta }{\Sigma }\right ) d\phi^2.
\label{e414}
\end{aligned}
\end{equation}

\section{Some properties of black holes in a dark matter halo}\label{s5}
Through the introduction of the previous sections, we have strictly given the metrics of the Schwarzschild-like black hole and the Kerr-like black hole in a dark matter halo. The difference between them and Schwarzschild black hole and Kerr black hole is that there is an additional dark matter term, that is $f(r), g(r)$ in (\ref{e38}) and (\ref{e412}). In this section, we will introduce the physical properties of this special black hole from several aspects such as horizons of black hole, shape of the ergosphere, motion equation of the neutral particle and so on. Finally, we compare these properties with the Kerr black hole, so that readers can understand the black hole in a dark matter halo more intuitively.
\subsection{Horizons of black hole in a dark matter halo} \label{s51}
The event horizon of a black hole is the outermost boundary of the black hole. In black hole physics, the event horizon of a black hole is a type of zero-surface with spacetime symmetry. General relativity shows that the Schwarzschild black hole has only one event horizon, which is determined entirely by the mass of the black hole. And the Kerr black hole has two event horizons, namely the event horizon and the Cauchy horizon, which are completely determined by the mass and angular momentum of the black hole.

\begin{figure*}[t!]
\centering
{
\includegraphics[width=0.31\columnwidth]{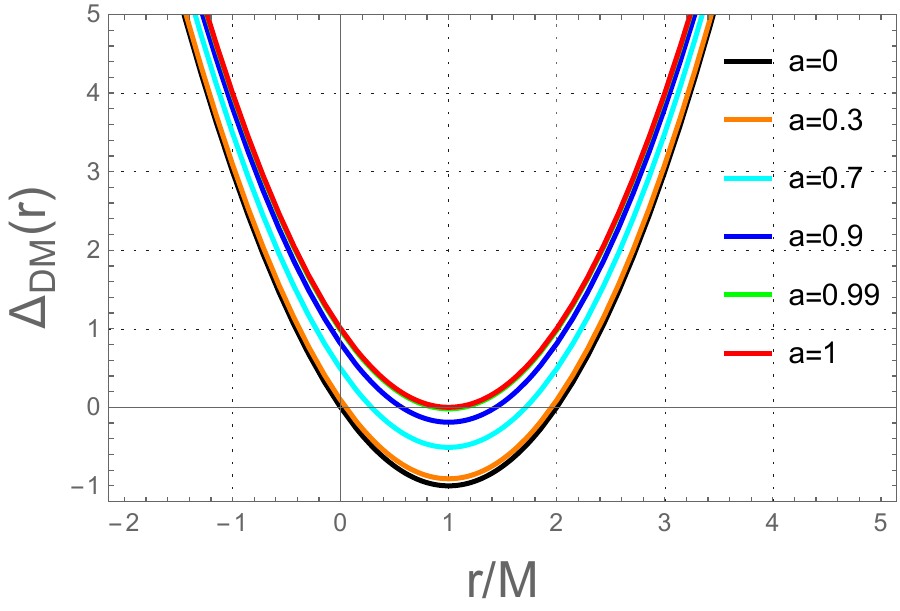}
}
{
\includegraphics[width=0.31\columnwidth]{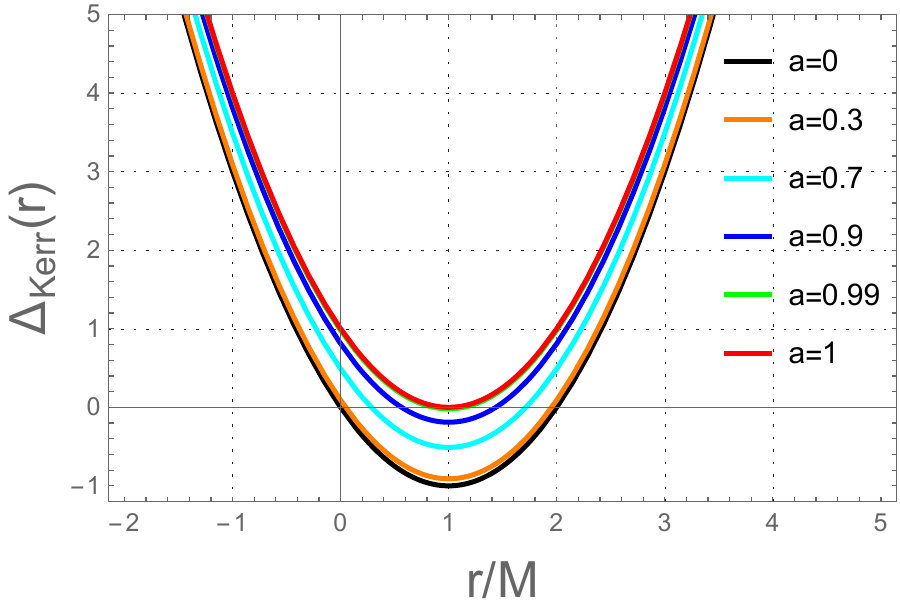}
}
{
\includegraphics[width=0.31\columnwidth]{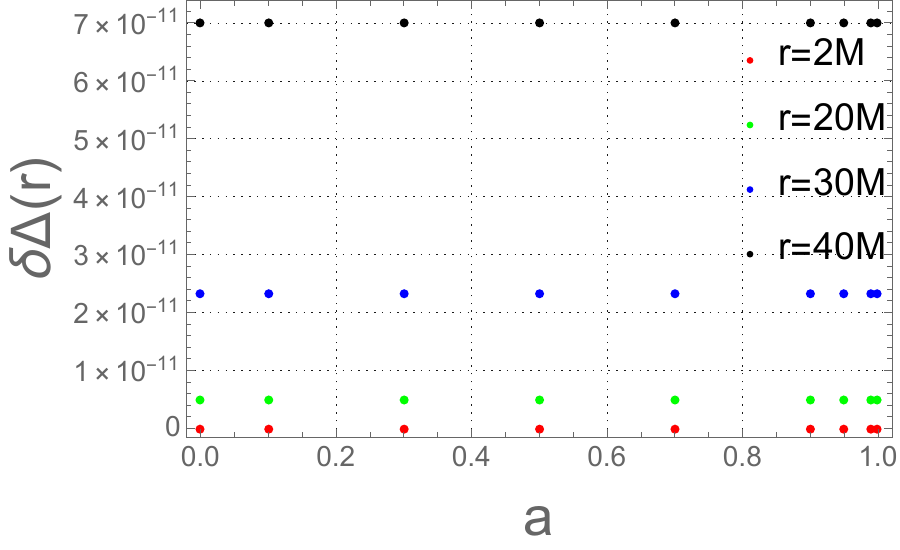}
}
\caption{The functional image of $\Delta$ as a function of independent variate $r$ in dark matter spacetime (left panel), Kerr spacetime (middle panel), respectively. The function image of $\Delta$ for different rotation parameters $a$ are given in the first two panels. The right panel is their comparison $\delta \Delta$ as a function of rotation parameter $a$ at the fixed $r$. The main calculation parameters are $M=1$, $\rho_\text{e}=6.9 \times 10^{6}$ $M_{\bigodot}/kpc^{3}$, $r_\text{e}=91.2$ $kpc$. We have converted these main calculation parameters to the black hole units before plotting.}
\label{f1}
\end{figure*}

For the Kerr-like black hole in a dark matter halo, its horizons can be found by solving $\Delta(r)=0$ in the formula (\ref{e413}), that is
\begin{equation}
\begin{aligned}
\exp\left ( \frac{2\alpha\text{ln}r-4\pi 2^{1-3/\alpha } e^{2/\alpha } r^2 ((r/r_\text{e})^{\alpha }/\alpha )^{-3/\alpha }\rho_\text{e} \Gamma [\frac{3}{\alpha },0,\frac{2(\frac{r}{r_\text{e}})^\alpha} {\alpha} ]}{\alpha}  \right )-2Mr+a^2=0,
\label{e51}
\end{aligned}
\end{equation}
In Figure \ref{f1}, we present the  functional image of $\Delta(r)$ as a function of the independent variate $r$ both in dark matter spacetime (left panel) and Kerr spacetime (middle panel). The number of these intersections with the x-axis in the image represents the number of roots of the equation $\Delta(r)=0$, which in black hole physics represents the horizons of the black hole. Our results show that black hole has the same number of horizons as Kerr black hole in a dark matter halo. The outer horizon decreases with the increase of the rotation parameter $a$, while the inner horizon increases with the increase of $a$ until the inner and outer event horizons are equal, that is, the extreme black hole. In order to distinguish the difference between the black hole in a dark matter halo and the Kerr spacetime, we define their difference $\delta \Delta(r)=\Delta_{\text{Kerr}}(r)-\Delta_{\text{DM}}(r)$, and then their differences as a function of rotation parameter $a$ are given in Figure \ref{f1} (right panel) at a fixed $r$. Our results show that the $\Delta(r)$ of the Kerr spacetime is greater than that of the dark matter halo, that is, $\Delta_{\text{Kerr}} > \Delta_{\text{DM}}$. In the right panel of Figure \ref{f1}, the maximum difference between them is approximately $7 \times 10^{-11}$, and the difference between them will increase with the increasing of variable parameter $r$. For example, when the rotation parameter $a=0.8$, the inner and outer horizons of the Kerr-like black hole in a dark matter halo are $r_\text{inner}=0.40000000000000002220, r_\text{outer}=1.6000000000000000888$, respectively, while the inner and outer horizons of the Kerr black hole are $r_\text{inner}=0.399999999999999996669 , r_\text{outer}=1.5999999999999998668$. This case shows that the inner and outer horizons of the Kerr-like black hole in a dark matter halo are slight larger than those of the Kerr black hole.

\subsection{Shape of ergosphere of black hole in a dark matter halo}
\begin{figure*}[t!]
\centering
{
\includegraphics[width=0.31\columnwidth]{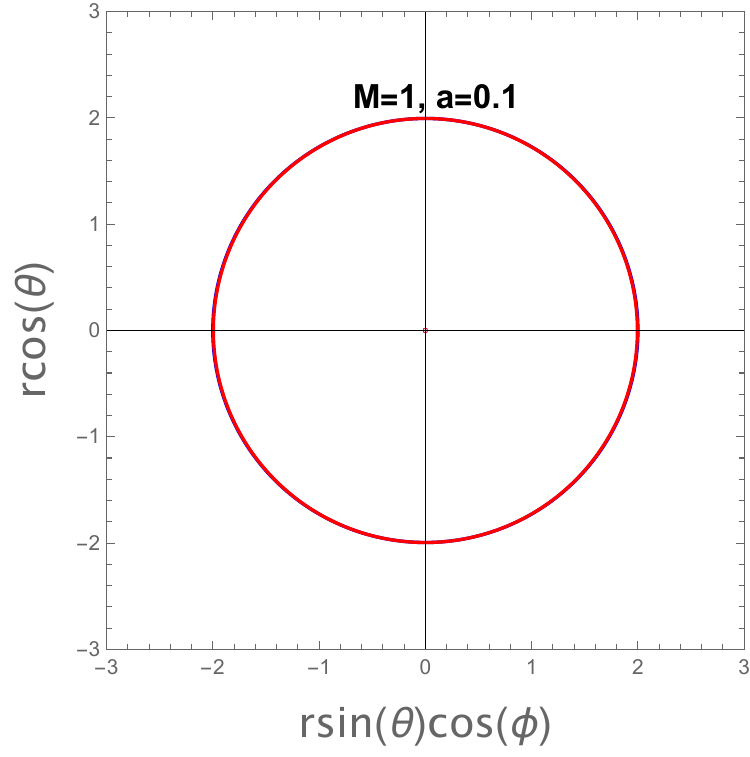}
}
{
\includegraphics[width=0.31\columnwidth]{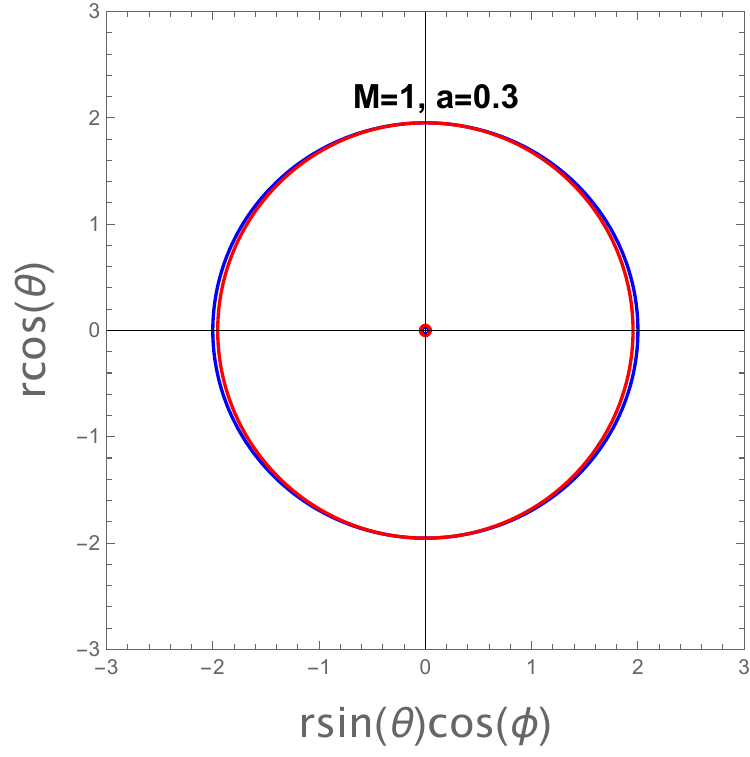}
}
{
\includegraphics[width=0.31\columnwidth]{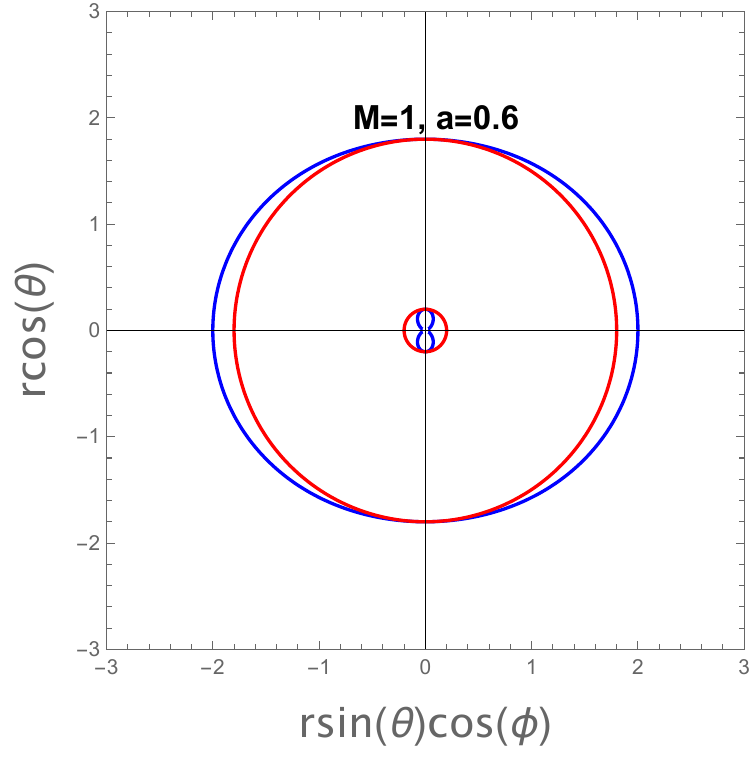}
}
{
\includegraphics[width=0.31\columnwidth]{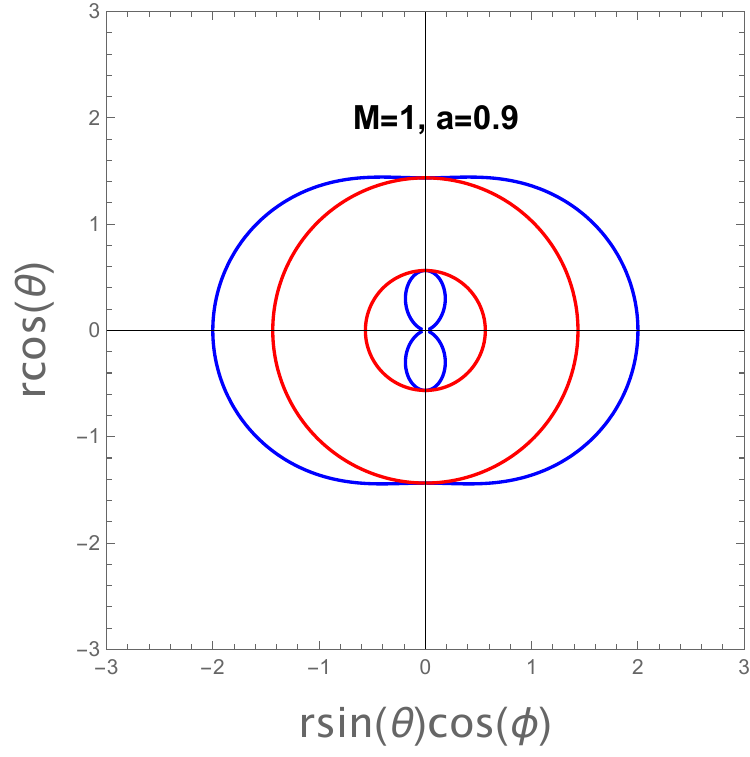}
}
{
\includegraphics[width=0.31\columnwidth]{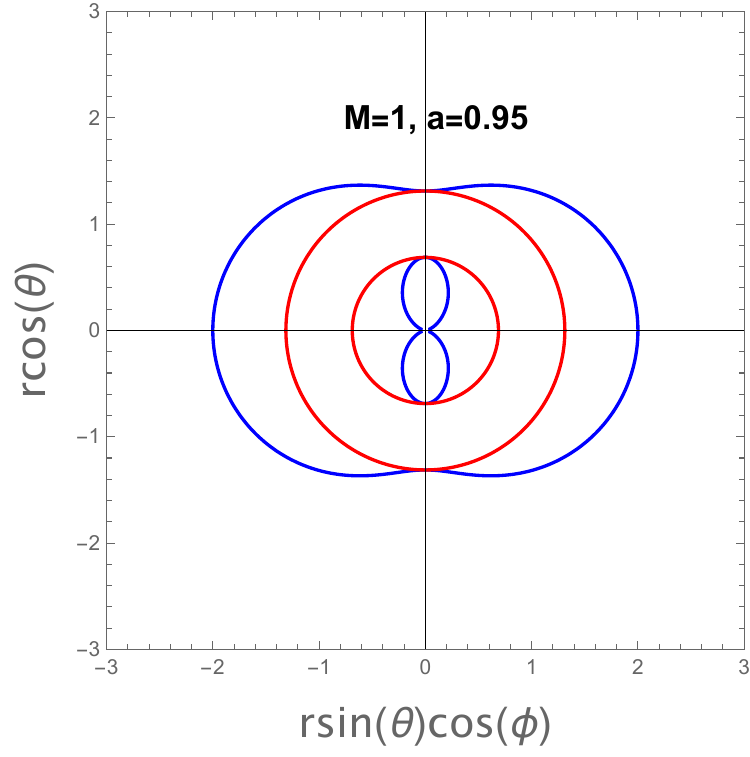}
}
{
\includegraphics[width=0.31\columnwidth]{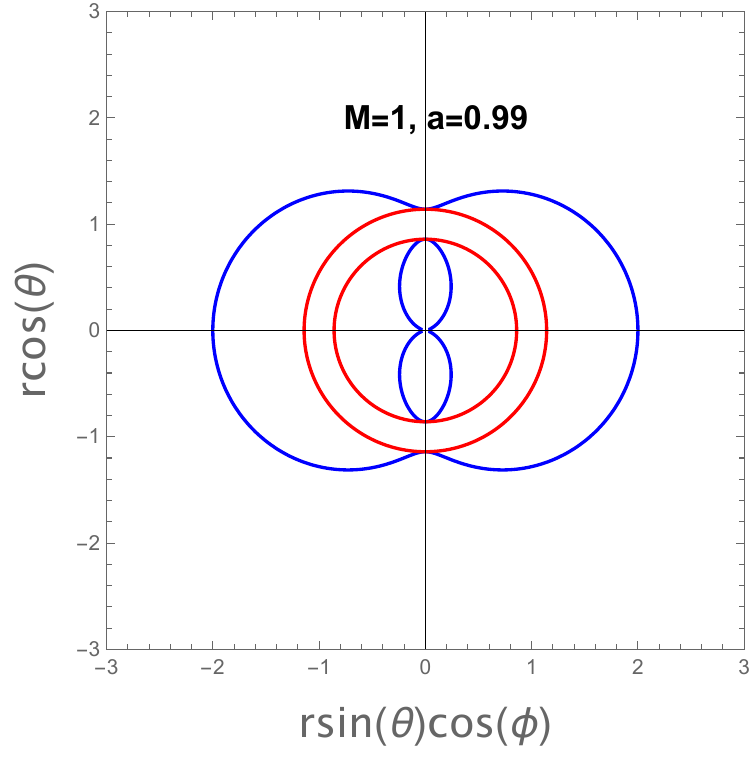}
}
\caption{The shape of ergosphere of the Kerr-like black hole in a dark matter halo with different rotation parameter $a$ in the $xz$-plane. The main calculation parameters are $M=1$, $\alpha=0.16$, $\rho_\text{e}=6.9 \times 10^{6}$ $M_{\bigodot}/kpc^{3}$, $r_\text{e}=91.2$ $kpc$. We have converted these main calculation parameters to the black hole units before plotting.}
\label{f2}
\end{figure*}

Next, let us study the shape of the ergosphere of this Kerr-like black hole in a dark matter halo. Please note that the ergosphere is only the outer region of the Kerr-like black hole. Firstly, in Figure \ref{f2}, we give the cross-section diagram of Kerr-like black hole in the dark matter halo in the $x z$-plane. In the black hole physics, the ergosphere is composed of two parts, one is the area surrounded by the outer infinite redshift surface and the outer event horizon, and the other is the area surrounded by the inner infinite redshift surface and the inner event horizon. The central area in the middle is a singular ring. Then, the inner and outer horizons and the infinite redshift surface of the Kerr-like black hole can be obtained by solving the Eqs. $\Delta(r)=0$ and $g_{t t}=0$, respectively, that is Eq. (\ref{e51}) and
\begin{equation}
\begin{aligned}
\exp\left ( \frac{2\alpha\text{ln}r-4\pi 2^{1-3/\alpha } e^{2/\alpha } r^2 ((r/r_\text{e})^{\alpha }/\alpha )^{-3/\alpha }\rho_\text{e} \Gamma [\frac{3}{\alpha },0,\frac{2(\frac{r}{r_\text{e}})^\alpha} {\alpha} ]}{\alpha}  \right )-2Mr+a^2\cos^2\theta=0,
\label{e52}
\end{aligned}
\end{equation}
From these figures in Figure \ref{f2}, for the Kerr-like black hole in a dark matter halo, we find that the outer horizon and outer infinite redshift surface decrease with the increase of rotation parameter $a$, while the inner horizon and inner infinite redshift surface increase with the increase of rotation parameter $a$. In addition, according to the calculation of the horizons in Figure \ref{f1}, dark matter may also increase the range of the ergosphere for the Kerr-like black hole.

\subsection{Motion equations of neutral particle near black hole in a dark matter halo}
In this subsection, we are interested in the neutral particle motion near the Kerr-like black hole in a dark matter halo, which can be solved by geodesic equations. Firstly, we rewrite the metric (\ref{e412}) in the Kerr-like form
\begin{equation}
\begin{aligned}
ds^2= & -(1-\frac{2R(r)r}{\Sigma }) dt^2 +\frac{\Sigma}{\Delta } dr^2 -\frac{4 R(r) r a \sin ^2 \theta }{\Sigma } dt d \phi + \Sigma d\theta^2 \\
&+\left ( (a^2+r^2)\sin^2 \theta +\frac{2 R(r) r a^2\sin^4\theta }{\Sigma }\right ) d\phi^2,
\label{e53}
\end{aligned}
\end{equation}
where,
\begin{equation}
\begin{aligned}
2R(r)=2M + r - r g(r), \quad  \Delta =r^2-2R(r)r+a^2, \quad \Sigma=r^2+a^2\cos^2 \theta
\label{e54}
\end{aligned}
\end{equation}
The geodesic equation of the neutral particle in a dark matter halo satisfy the Lagrangian equation \cite{Khan:2019gco},
\begin{equation}
\mathscr{L}=\frac{1}{2}g_{{\mu \nu }}\dot{x}^\mu \dot{x}^\nu,
\label{e55}
\end{equation}
where, $g_{\mu\nu}$ is the metric component of (\ref{e53}) and $\dot{x}^\mu$ denotes the first-order partial derivative to the affine parameter of the coordinate $x^\mu$. The 4-momentum $p_\mu$ of the neutral particle from (\ref{e53}) is $p_\mu =\partial \mathscr{L} / \partial \dot{x}^\mu =g_{\mu\nu}\dot{x}^\nu$, and then
\begin{equation}
\begin{aligned}
-p_t=g_{tt}\dot{t}+g_{t\phi }\dot{\phi }=E, \quad p_\phi=g_{t\phi }\dot{t}+g_{\phi\phi }\dot{\phi }=L, \quad p_r=g_{rr}\dot{r}, \quad p_\theta =g_{\theta \theta }\dot{\theta },
\label{e56}
\end{aligned}
\end{equation}
where, $E$ and $L$ are the energy and angular momentum of the neutral particle, respectively. In addition, $p_t$ and $p_\phi$ are a conserved quantity because the Lagrangian is not related to the coordinates $t$ and $\phi$. Therefore, the Kerr-like black hole in a dark matter halo has the properties of steady state and axisymmetric. Through the first two equations in (\ref{e56}), we can reverse solve $\dot{t}$ and $\dot{\phi}$, that is
\begin{equation}
\begin{aligned}
\dot{t}&=\frac{1}{\Sigma \Delta }(-E(a^2+r^2)\Sigma +2ar(-L-aE\sin^2\theta)R(r)),\\
\dot{\phi}&=-\frac{1}{\Sigma \Delta \sin^2\theta }(-L \Sigma +2r(L+aE\sin^2\theta)R(r)).\\
\label{e57}
\end{aligned}
\end{equation}
Based on the momentum and Lagrangian above, the Hamiltonian of the neutral particle motion can be written as
\begin{equation}
\begin{aligned}
H=-p_t\dot{t}+p_r\dot{r}+p_\theta\dot{\theta}+p_\phi\dot{\phi}-\mathscr{L},
\label{e58}
\end{aligned}
\end{equation}
Taking (\ref{e53}), (\ref{e55}) and (\ref{e56}) into account, (\ref{e58}) can be written as
\begin{equation}
\begin{aligned}
2H&=-(g_{tt}\dot{t}+g_{t\phi}\dot{\phi})\dot{t}+g_{rr}\dot{r}^2+g_{\theta\theta}\dot{\theta}^2+(g_{t\phi}\dot{t}+g_{\phi\phi}\dot{\phi})\dot{\phi}\\
&=E\dot{t}+L\dot{\phi}+\frac{\Sigma }{\Delta }\dot{r}^2+\Sigma \dot{\theta}^2=m=constant,
\label{e59}
\end{aligned}
\end{equation}
where, $m$ is the mass of the neutral particle, and $m=0$ corresponds to the motion equation of photons. The results (\ref{e57}), (\ref{e58}), (\ref{e59}) are very important, because they can be used to study the properties of particle near black hole. Eq. (\ref{e59}) is a second order partial differential equation. Usually, a separation constant can be introduced to reduce it to a radial equation and an angular equation. Where, the radial equation describes the motion of the neutral particle near the black hole. As an example, we study the neutral particle whose orbits lie in the equatorial plane (i.e., $\theta=\pi/2, m=0$), and its motion equation is given by
\begin{equation}
\begin{aligned}
\dot{r}^2=E^2+\frac{1}{r^2}(a^2E^2-L^2)+\frac{2R(r)}{r^3}(aE+L)^2.
\label{e510}
\end{aligned}
\end{equation}
In particular, when the dark matter is absent ($\rho_\text{e}=0$), that is $g(r) = 1$ in $R(r)$, the radial equation will return to the case of the Kerr vacuum.

\subsection{The geodesic equations of lightlike particle and photon regions}
In the previous subsection, we calculated the motion equation for the neutral particle using the momentum and Lagrangian based on the metric (\ref{e53}). Here, we rewrite this metric as
\begin{equation}
\begin{aligned}
ds^2= & -(1-\frac{2R(r)r}{\Sigma }) dt^2 +\frac{\Sigma}{\Delta } dr^2 -\frac{4 R(r) r a \sin ^2 \theta }{\Sigma } dt d \phi + \Sigma d\theta^2 \\
&+ \frac{\sin^2\theta}{\Sigma }((r^2+a^2)^2-a^2\Delta \sin^2\theta) d\phi^2,
\label{e511}
\end{aligned}
\end{equation}
where,
\begin{equation}
\begin{aligned}
2R(r)=2M + r -r g(r), \quad  \Delta =r^2-2R(r)r+a^2, \quad \Sigma=r^2+a^2\cos^2 \theta.
\label{e512}
\end{aligned}
\end{equation}
In this subsection, we will focus on the motion of light particles and their photon regions. Therefore, we need to know the geodesic equations describing the lightlike particles. The lightlike geodesic equation can be obtained by combining the momentum and the Hamilton-Jacobi equation
\begin{equation}
-\frac{\partial S}{\partial \sigma }=\frac{1}{2}g^{\mu\nu}\frac{\partial S}{\partial x^\mu}\frac{\partial S}{\partial x^\nu},
\label{e513}
\end{equation}
where, $S$ is Hamiltonian-Jacobi action, $\sigma$ is an affine parameter, $x^\mu$ is the coordinate component and $g^{\mu\nu}$ is the inverse metric of (\ref{e511}). For a test particle moving along the geodesic of the Kerr-like black hole, the Hamiltonian-Jacobi action $S$ can always be written in the separated form,
\begin{equation}
S=\frac{1}{2}m^2\sigma -Et+L\phi +S_\theta (\theta )+S_r(r),
\label{e514}
\end{equation}
where, $m$ is the mass of the particle (here the photon takes $m=0$). Taking (\ref{e511}) and (\ref{e514}) into account, the Eq. (\ref{e513}) can be divided into the following two equations,
that is
\begin{equation}
\Delta  S_r'(r)^2-\frac{((r^2+a^2)E-aL)^2}{\Delta}+(L-aE)^2=-K,
\label{e515}
\end{equation}
and
\begin{equation}
S_\theta'(\theta )^2-a^2 E^2 \cos^2\theta+L^2 \cot^2\theta=K,
\label{e516}
\end{equation}
where, $S_r'(r)$ denotes $\partial S_r(r)/\partial r$, $S_\theta'(\theta )$ denotes $\partial S_\theta(\theta)/\partial \theta$ and $K$ is a constant obtained by separating variables, which is called Carter constant \cite{Carter:1968rr}. Besides, we note that $\partial S /\partial x^\mu = p_\mu$, and then it is not difficult to obtain the geodetic equations of the lightlike particles
\begin{equation}
\begin{aligned}
&\Sigma  \dot{r} = \sqrt{\mathcal{R}},\\
&\Sigma  \dot{\theta } = \sqrt{\Theta},\\
&\Sigma \dot{t}=E\left ( \frac{(a^2+r^2)(a^2+r^2-a\lambda )}{\Delta }-a( a \sin^2\theta - \lambda) \right ),\\
&\Sigma \dot{\phi }=-E\left ( \frac{a(a^2+r^2+a \lambda \csc^2 \theta)}{\Delta }-(a+\lambda \csc^2\theta ) \right ),\\
\label{e517}
\end{aligned}
\end{equation}
where,
\begin{equation}
\begin{aligned}
\lambda &=L/E, \\
\eta &=K/E^2,\\
\mathcal{R}&=E^2\left ( (a^2+r^2-a\lambda )^2- \Delta ((\lambda-a)^2+\eta ) \right ), \\
\Theta &=E^2\left (\eta -\lambda ^2\cot^2\theta +a^2\cos^2\theta \right ),
\label{e518}
\end{aligned}
\end{equation}
the dot ($\cdot $) in the Eq. (\ref{e517}) denotes the derivative to the affine parameter $\sigma$, that is $d/d\sigma$. Next, let's study and analyze the motion of photons in circular orbits. The motion of a photon usually satisfies the radial equation, that is the first equation of Eq. (\ref{e517}), and it can be rewritten as 
\begin{equation}
\begin{aligned}
\left ( \Sigma \frac{dr}{d\sigma } \right )^2+V_\text{eff}(r)=0,
\end{aligned}
\end{equation}
where, the effective potential is
\begin{equation}
\begin{aligned}
V_\text{eff}(r)/E^2=-(a^2+r^2-a\lambda )^2 + \Delta ((\lambda-a)^2+\eta ),
\label{ne520}
\end{aligned}
\end{equation}
where, the function $ \Delta$ can be obtained in (\ref{e512}). For a circular photon orbit, it satisfies the following condition
\begin{equation}
\begin{aligned}
V_\text{eff}(r)\mid _{r=r_\text{c}}=0, \quad \frac{\partial V_\text{eff}(r)}{\partial r}\mid _{r=r_\text{c}}=0,
\label{e519}
\end{aligned}
\end{equation}
where, $r_\text{c}$ is the orbital radius of photon sphere and then we can obtain
\begin{equation}
\begin{aligned}
\lambda &=\frac{-4r\Delta(r)+a^2\Delta '(r)+r^2\Delta '(r)}{a \Delta '(r)}\mid _{r=r_\text{c}}, \\
\eta &=\frac{r^2 (16a^2\Delta (r)-16\Delta (r)^2+8r\Delta (r)\Delta '(r)-r^2\Delta '(r)^2)}{a^2\Delta '(r)^2}\mid _{r=r_\text{c}},
\label{e520}
\end{aligned}
\end{equation}
where, $\Delta '(r)$ represents the derivative of $\Delta$ with respect to $r$. Taking (\ref{e520}) into the third equation of (\ref{e518}) can be obtained
\begin{equation}
V_{\text{eff}}''(r)=8 E^2 \left(r^2+\frac{2 r\Delta (r) \left(\Delta '(r)-r \Delta ''(r)\right)}{\Delta'(r)^2}\right) \mid _{r=r_\text{c}},
\label{e521}
\end{equation}
In particular, if $V_\text{eff}''(r_\text{c})>0$, the radial equation corresponding to the lightlike geodesics is unstable, but $V_\text{eff}''(r_\text{c})<0$ is stable. The unstable orbit of the photon will determine the shape of shadow of the black hole. The range of the photon regions $r_\text{c}$ can be obtained by the equation $\Theta \geq 0$ with case of $\theta=\pi/2$, and then taking (\ref{e520}) into the last equation of (\ref{e518}) and then we can obtain the following form
\begin{equation}
\begin{aligned}
16\Delta(r)^2 + r^2\Delta'(r)^2 - 8\Delta(r)(2a^2+r\Delta'(r)) \mid _{r=r_\text{c}} &\leq 0,
\label{e522}
\end{aligned}
\end{equation}
From Eq. (\ref{e522}), once the parameters of the black hole are determined, the range of photon region $r_\text{c}$ may be $r_{\text{cmin}}\leq r_\text{c} \leq r_{\text{cmax}}$. Here, $r_{\text{cmin}}, r_{\text{cmax}}$ are the minimum and maximum positions of the photon regions, respectively. It can be seen that there are maximum and minimum values of the photon sphere.

\begin{figure*}[t!]
\centering
{
\includegraphics[width=0.45\columnwidth]{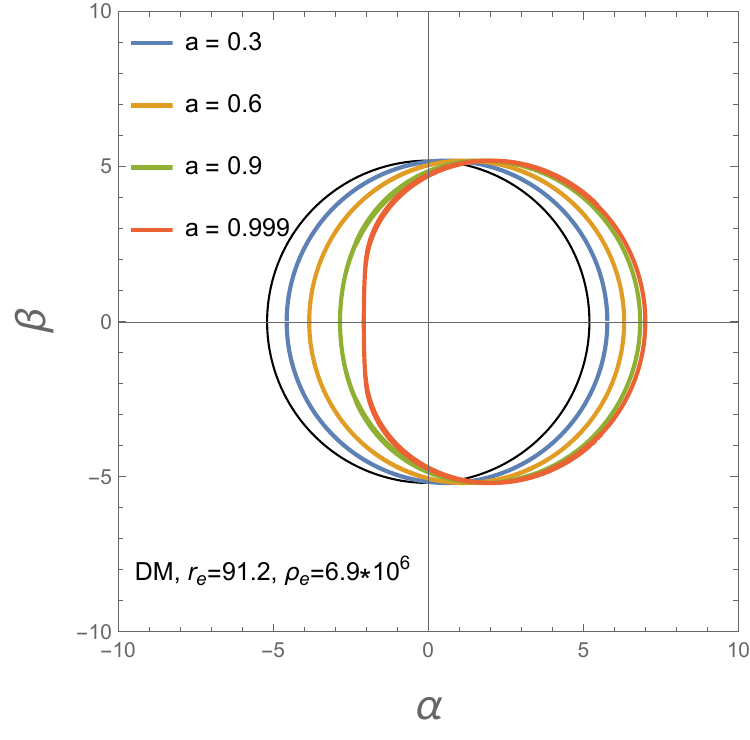}
}
{
\includegraphics[width=0.45\columnwidth]{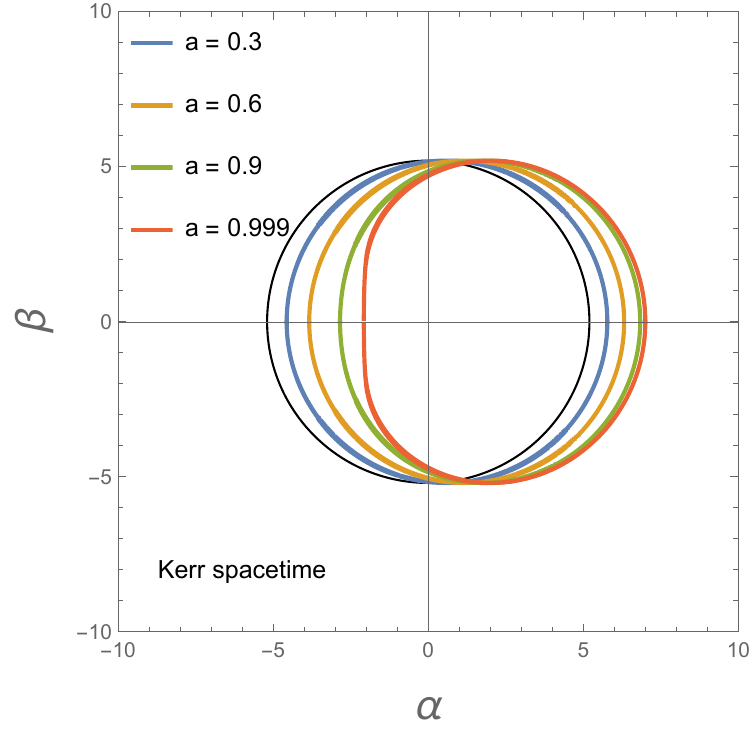}
}
\caption{The shape of shadow of the Kerr-like black hole and the Kerr black hole both with the different rotation parameter $a$ for observers located at equatorial plane. The black circle in the center of the coordinate axis represents the shadow of the Schwarzschild black hole. The main calculation parameters are $M=1$, $\alpha=0.16$, $\rho_\text{e}=6.9 \times 10^{6}$ $M_{\bigodot}/kpc^{3}$, $r_\text{e}=91.2$ $kpc$. We have converted these main calculation parameters to the black hole units before plotting.}
\label{f3}
\end{figure*}
\begin{figure*}[t!]
\centering
{
\includegraphics[width=0.31\columnwidth]{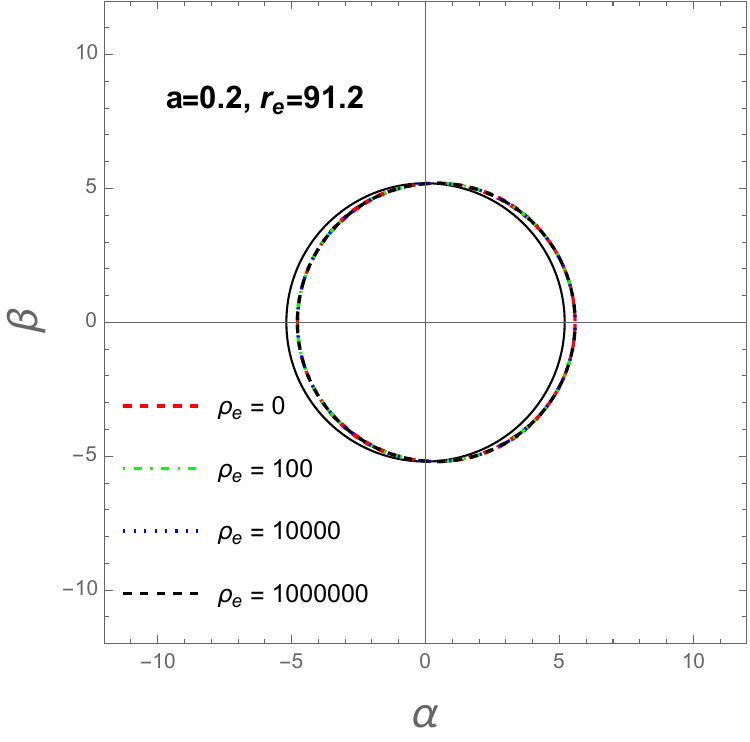}
}
{
\includegraphics[width=0.31\columnwidth]{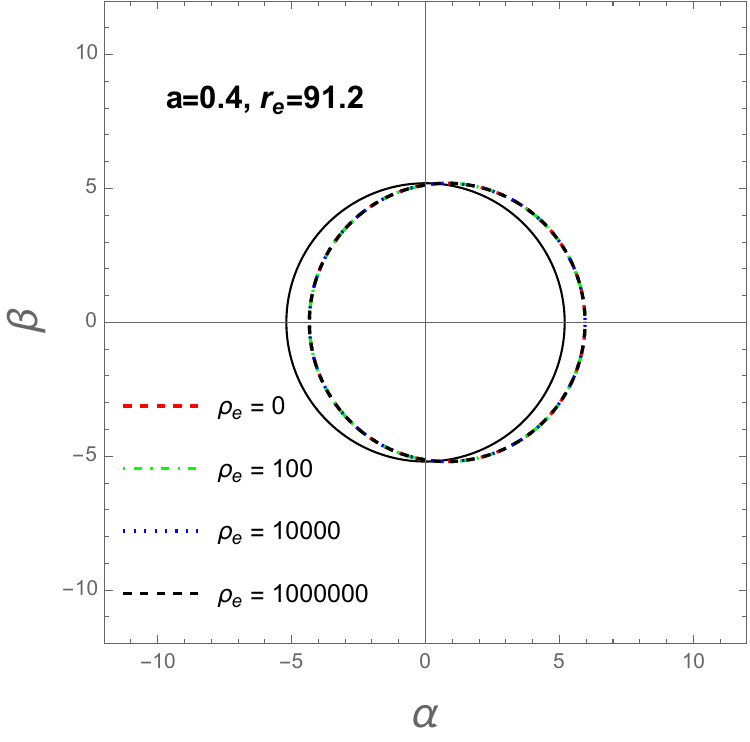}
}
{
\includegraphics[width=0.31\columnwidth]{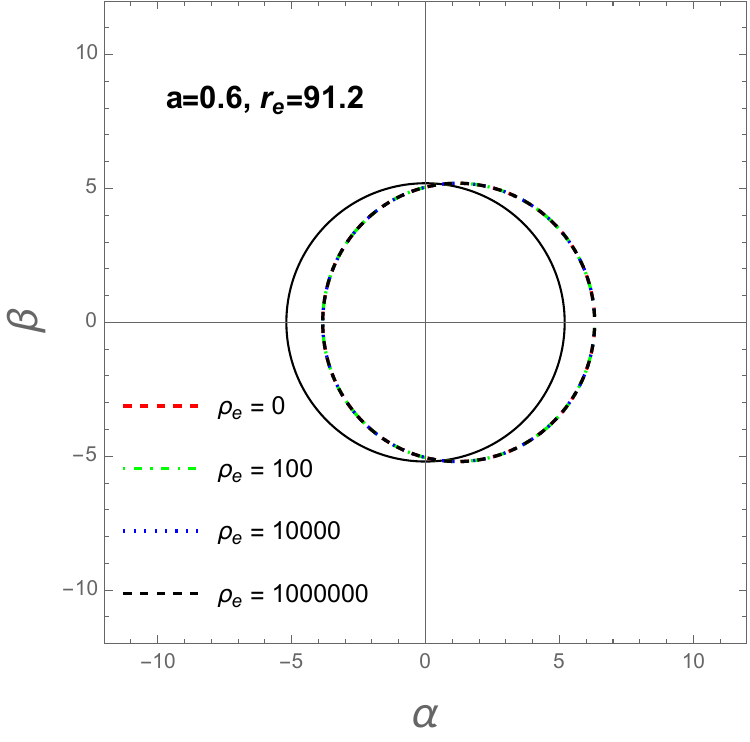}
}
{
\includegraphics[width=0.31\columnwidth]{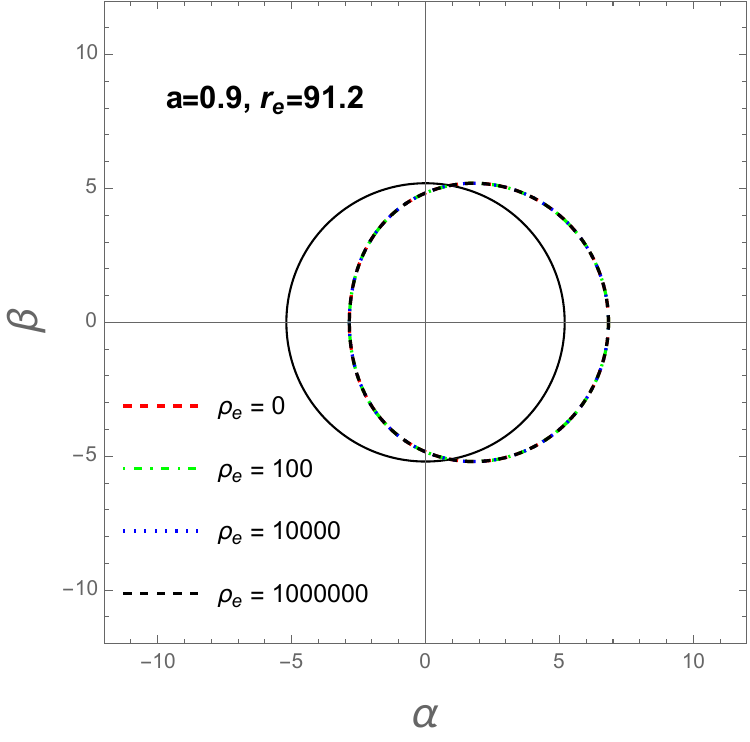}
}
{
\includegraphics[width=0.31\columnwidth]{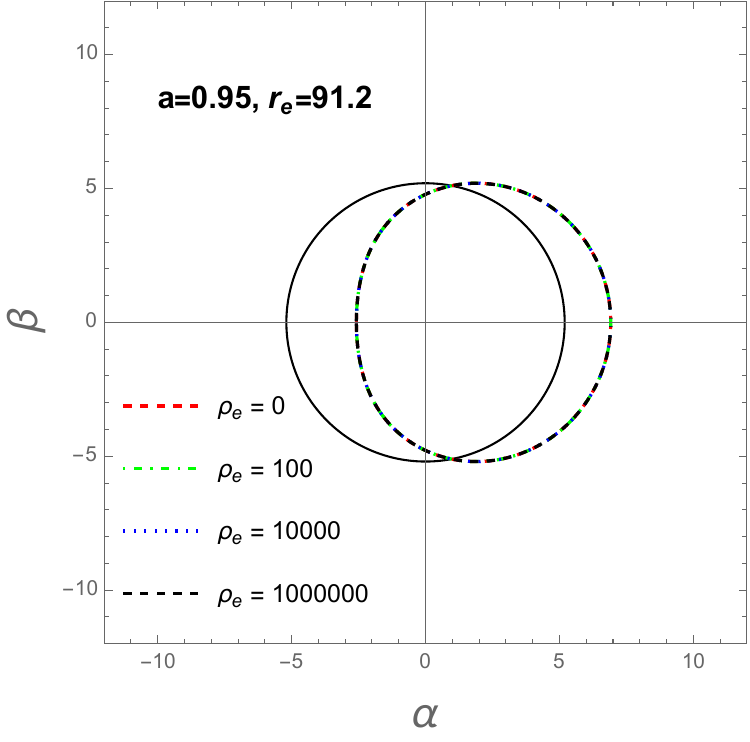}
}
{
\includegraphics[width=0.31\columnwidth]{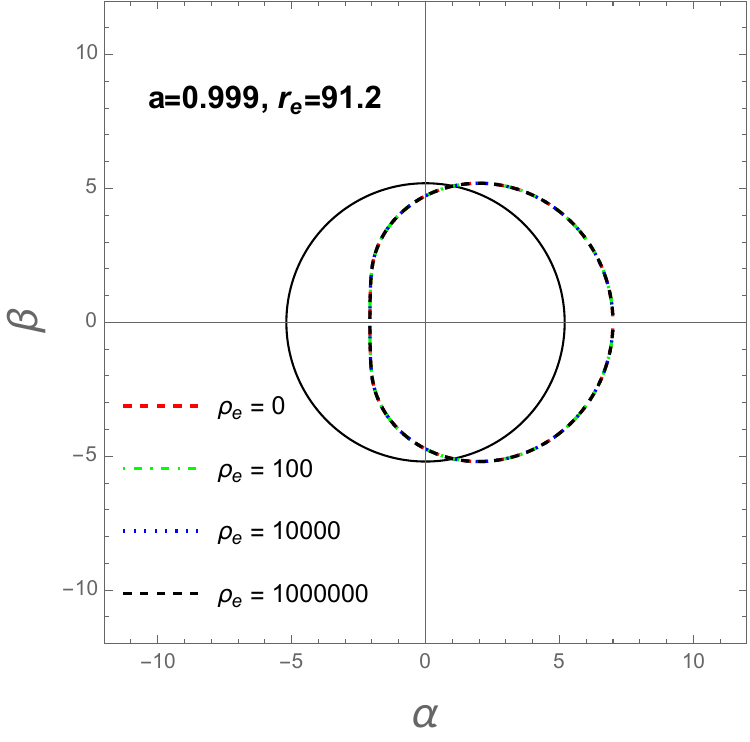}
}
\caption{The shape of shadow of the Kerr-like black hole with the different dark matter densities $\rho_\text{e} (M_\odot/kpc^3)$ for observers located at equatorial plane. The black circle in the center of the coordinate axis represents the shadow of the Schwarzschild black hole. The main calculation parameters are $M=1$, $\alpha=0.16$, $r_\text{e}=91.2$ $kpc$. We have converted these main calculation parameters to the black hole units before plotting.}
\label{f4}
\end{figure*}

\subsection{The shadow of the Kerr-like black hole in a dark matter halo}
In the previous subsection, we calculated the geodesic equations for the motion of lightlike particle and the photon regions in a dark matter halo. Next, we will study the black hole shadow of the Kerr-like black hole in a dark matter halo based on these geodesic equations, and compare them with that of the Kerr black hole. For an asymptotically flat spacetime, we can easily calculate the shadow of the black hole in celestial coordinates, and the formula of the corner radius of the shadow has the same form as that of the shadow corner radius of the black hole in Minkowski spacetime \cite{Hioki:2009na}, that is
\begin{equation}
\begin{aligned}
\alpha &=\lim_{r_0\rightarrow \infty }\left ( -r_0^2 \sin(\theta )\frac{d\phi }{dr}\mid _{\theta =\theta _0} \right ),\\
\beta &=\lim_{r_0\rightarrow \infty }\left [ r_0^2\frac{d\theta }{dr}\mid _{\theta =\theta _0} \right ],
\label{e523}
\end{aligned}
\end{equation}
where, $(r_0,\theta_0)$ is the position of the observer, and the motion of the light particle is described by $d\phi /dr $ and $d\theta /dr$ from Eqs. (\ref{e517}). Here, we choose to observe the shape of shadow of the Kerr-like black hole on the equator plane, that is, $\theta_0=\pi/2$. Observed in the equatorial plane, the shadow of Kerr-like black hole is the outgoing light perpendicular to the rotation axis of the black hole received by an observer at a certain distance. For more details on the introduction of black hole shadow, we can refer to this brief review in Ref. \cite{Perlick:2021aok}. This viewing angle can provide some interesting and important information, because from this angle, we can see the impacts of the rotation parameter $a$ of the black hole on the shadow of the black hole. Then the above equation can be written as
\begin{equation}
\begin{aligned}
\alpha &=-\frac{\lambda }{\sin(\theta )}\mid_{\theta =\frac{\pi}{2}}=-\lambda,  \\
\beta &=\pm \sqrt{\eta -a^2\cos(\theta )^2-\lambda ^2\cot(\theta )^2}\mid_{\theta =\frac{\pi}{2}}=\pm\sqrt{\eta }.
\label{e524}
\end{aligned}
\end{equation}
Taking (\ref{e413}) and (\ref{e520}) into account, we can obtain the Celestial coordinates in a dark matter halo, that is
\begin{equation}
\begin{aligned}
\alpha =&-\frac{r^2 \left(a^2+r^2\right) g'(r)+2 r \left(a^2-r^2\right) g(r)-2 a^2 M-4 a^2 r+6 M r^2}{a \left(r^2 \left(-g'(r)\right)-2 r g(r)+2 M\right)},\\
\beta =&\pm \sqrt{\frac{r^5 \left(8 a^2 g'(r)+\left(2 g(r)-r g'(r)\right) \left(r^2 g'(r)-2 r g(r)+12
   M\right)\right)+4 M r^3 \left(4 a^2-9 M r\right)}{a^2 \left(r^2 g'(r)+2 r g(r)-2 M\right)^2}}
\label{e527}
\end{aligned}
\end{equation}
where, $g'(r)$ denotes $d g(r)/dr$ and $g(r)=f(r)$ can be obtained in (\ref{e26}) in this work. In the Figure \ref{f3}, we give the shadow of the Kerr-like black hole and the Kerr black hole with the different rotation parameter $a$ for observers located at equatorial plane. At the same time, we also give the shadow of the non-rotating black hole, that is, the Schwarzschild black hole, and the critical value of its shadow is $r_\text{sch}=3\sqrt{3}M_{\text{BH}}$ \cite{Misner:1973prb,Perlick:2021aok}. Our results show that with the increasing of the rotation parameter $a$, the degree of distortion of the black hole shadow gradually increases. Besides, in Figure \ref{f4}, we also study the impact of different dark matter parameters on the black hole shadow, and compare them with the Kerr black hole ($\rho_\text{e}=0$). We find that the shape of the black hole shadow in a dark matter halo is similar to the Kerr black hole, which also may show that the dark matter has little impact on the shape of the black hole shadow. On the other hand, the results in Figure \ref{f1} show that the horizons of a black hole in a dark matter halo is slightly larger than the GR black hole, but the difference between them is very small, which is a challenge for astronomical observation. But what may be certain is that the shadow of a dark matter black hole is slightly larger than that of a GR black hole. Therefore, when plotting Figures \ref{f3} and \ref{f4}, in order to ensure the integrity of the shadows, we did not show these small differences as the right panel of Figure \ref{f1} like. In the next subsection of discussion, we can quantitatively analyze these differences between them from another perspective.

\begin{figure*}[t!]
\centering
{
\hspace{-2cm}\includegraphics[width=0.5 \columnwidth]{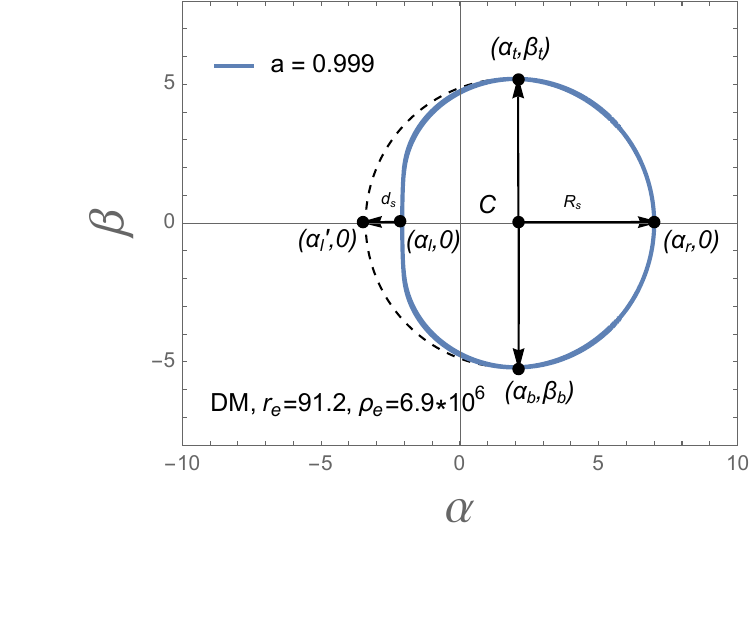}
}
\caption{The coordinate diagram of the Kerr-like black hole shadow $(a=0.999)$ in a dark matter halo. Here, we define the radius $R_s$, the distortion parameter $\delta_s$ of the shadow in the Kerr-like black hole and the coordinates of some specific points. The main calculation parameters are $M=1$,$\alpha=0.16$, $\rho_\text{e}=6.9 \times 10^{6}$ $M_{\bigodot}/kpc^{3}$, $r_\text{e}=91.2$ $kpc$. We have converted these main calculation parameters to the black hole units before plotting.}
\label{f5}
\end{figure*}

\begin{figure*}[t!]
\centering
{
\includegraphics[width=0.31\columnwidth]{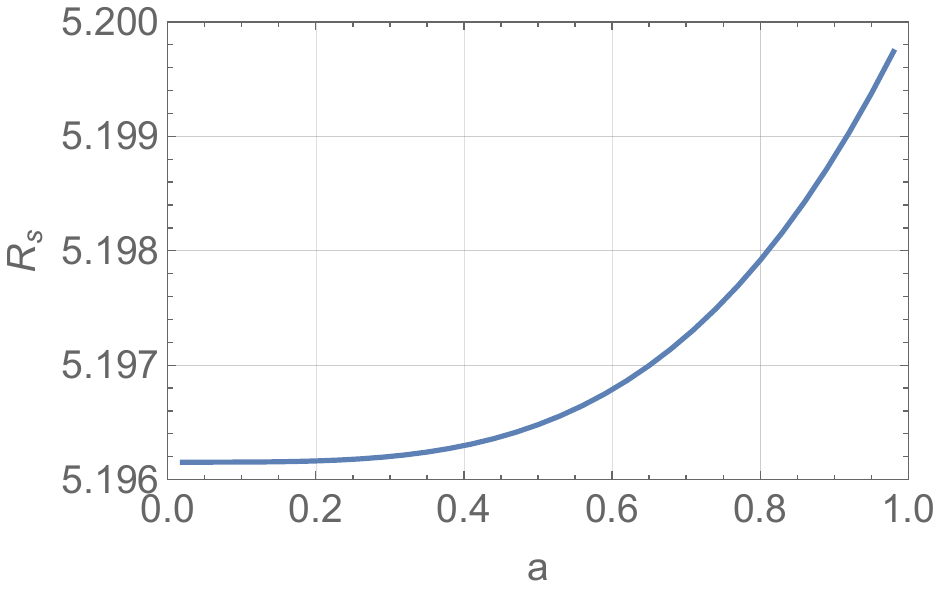}
}
{
\includegraphics[width=0.31\columnwidth]{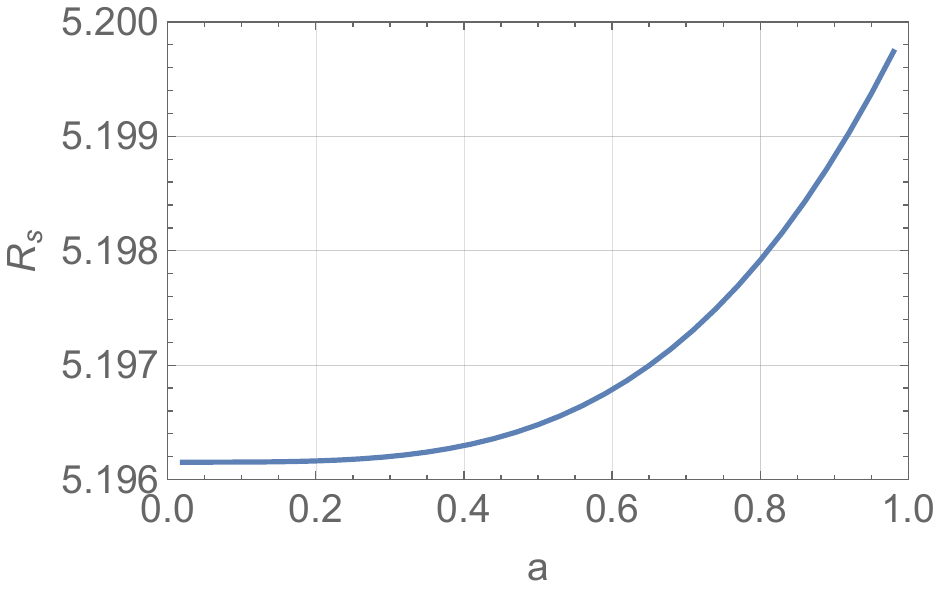}
}
{
\includegraphics[width=0.31\columnwidth]{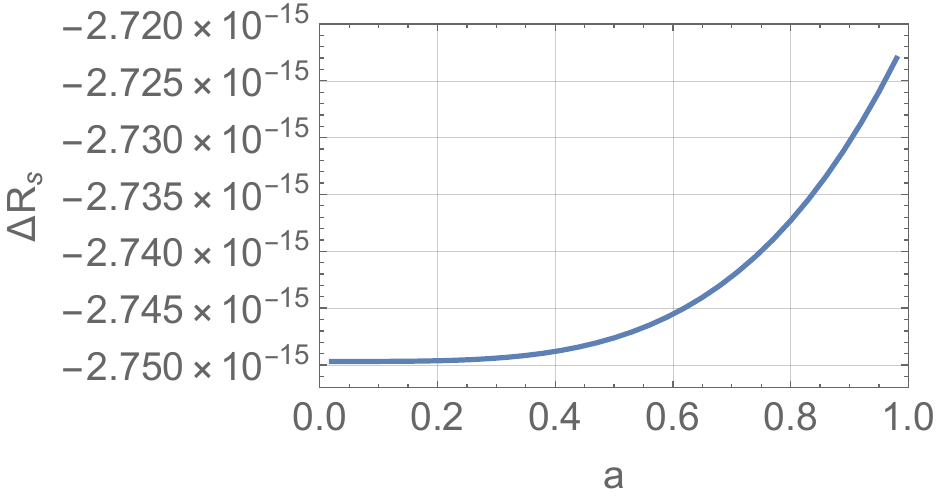}
}
{
\includegraphics[width=0.31\columnwidth]{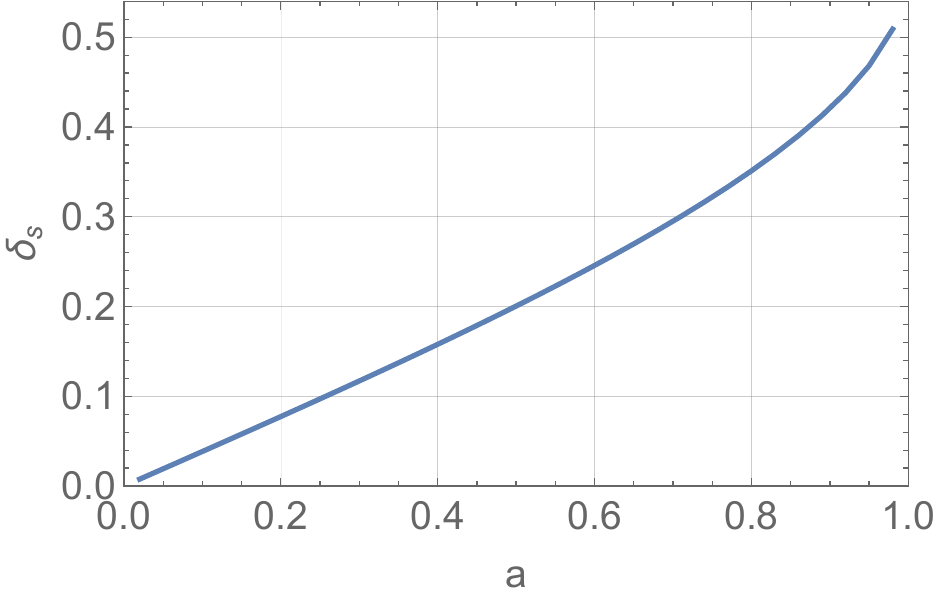}
}
{
\includegraphics[width=0.31\columnwidth]{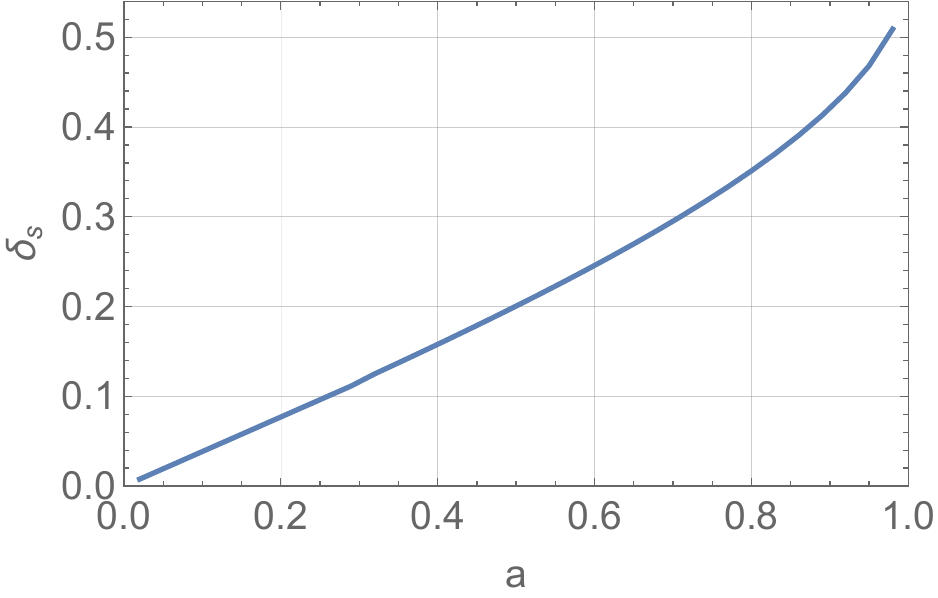}
}
{
\includegraphics[width=0.31\columnwidth]{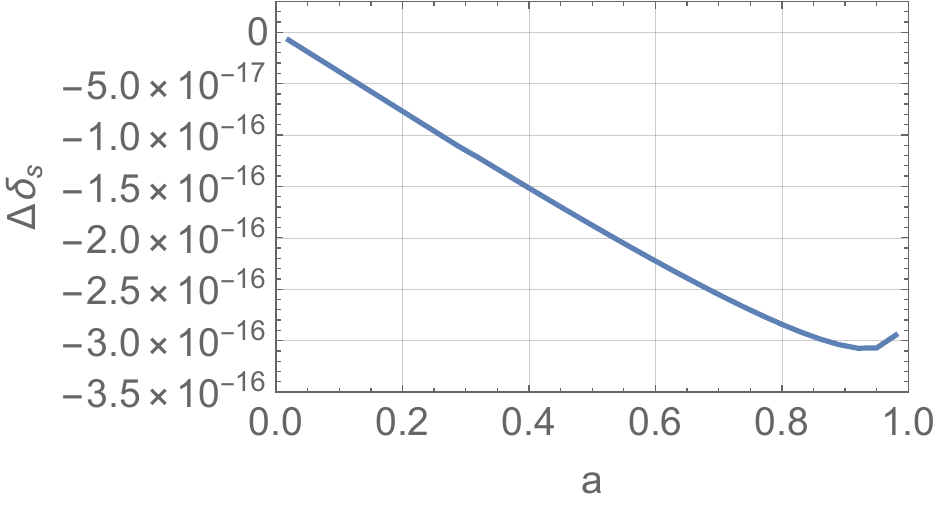}
}
{
\includegraphics[width=0.31\columnwidth]{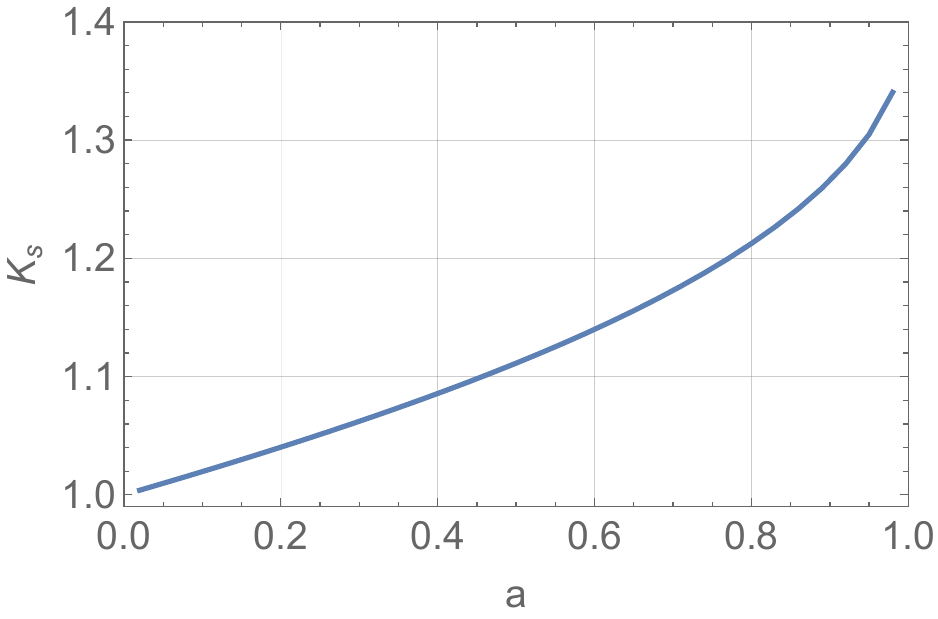}
}
{
\includegraphics[width=0.31\columnwidth]{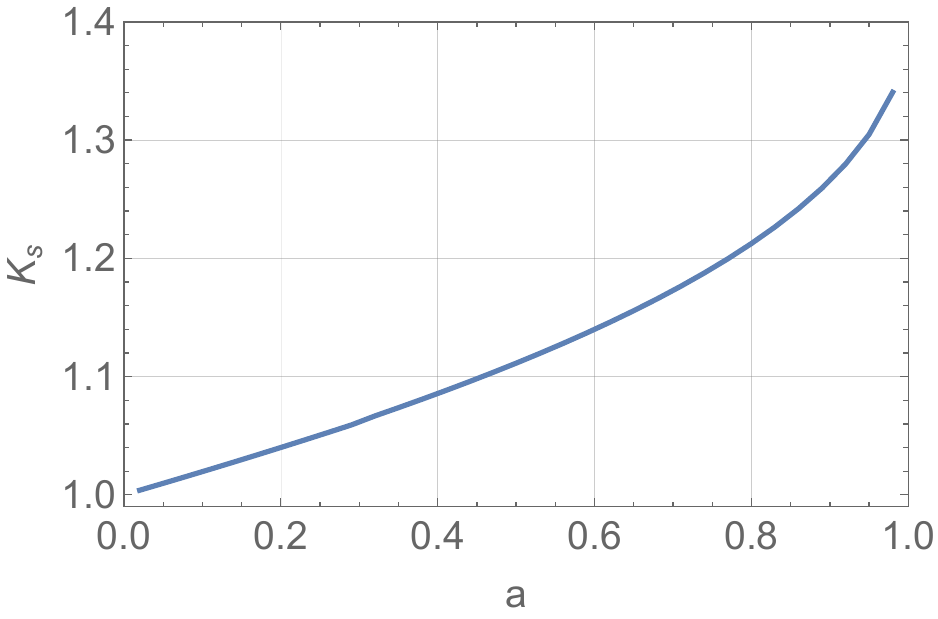}
}
{
\includegraphics[width=0.31\columnwidth]{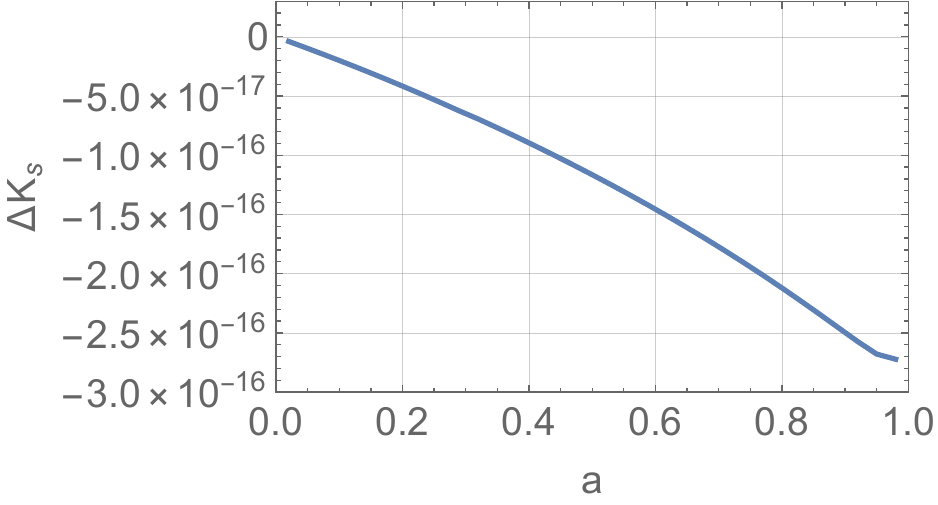}
}
\caption{The observables $R_\text{s}$, $\delta_\text{s}$ and $K_\text{s}$ separately as a function of the rotation parameter $a$ of the Kerr spacetime (left panels) and the Kerr-like black hole (middle panels) in the galactic center of M87. The right panels are the differences of these observables between Kerr black hole and Kerr-like black hole. The main calculation parameters are $M=1$, $\alpha=0.16$, $\rho_\text{e}=6.9 \times 10^{6}$ $M_{\bigodot}/kpc^{3}$, $r_\text{e}=91.2$ $kpc$  and the step size of the rotation parameter $a$ is $0.03$ from $0$ to $1$. We have converted these main calculation parameters to the black hole units before plotting.}
\label{f6}
\end{figure*}

\begin{figure*}[t!]
\centering
{
\includegraphics[width=0.31\columnwidth]{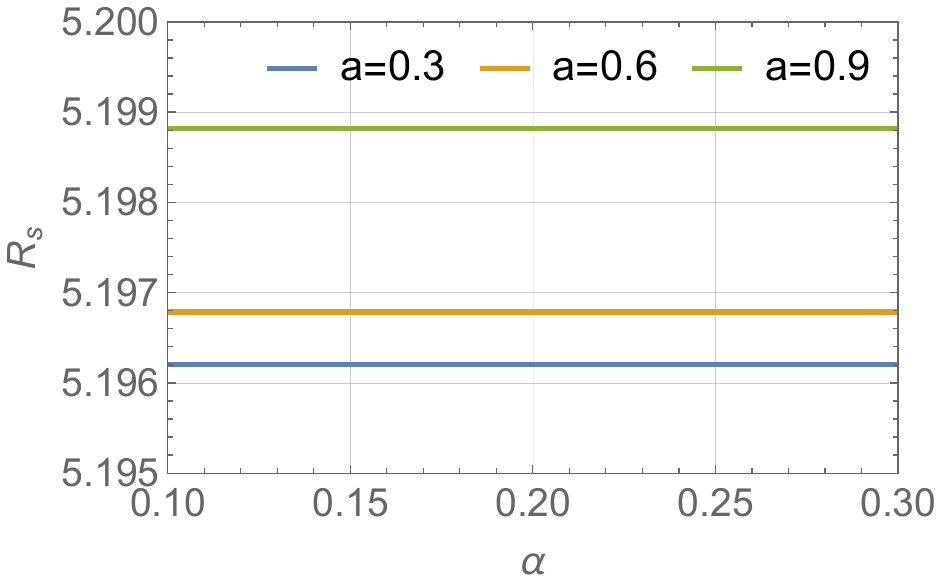}
}
{
\includegraphics[width=0.31\columnwidth]{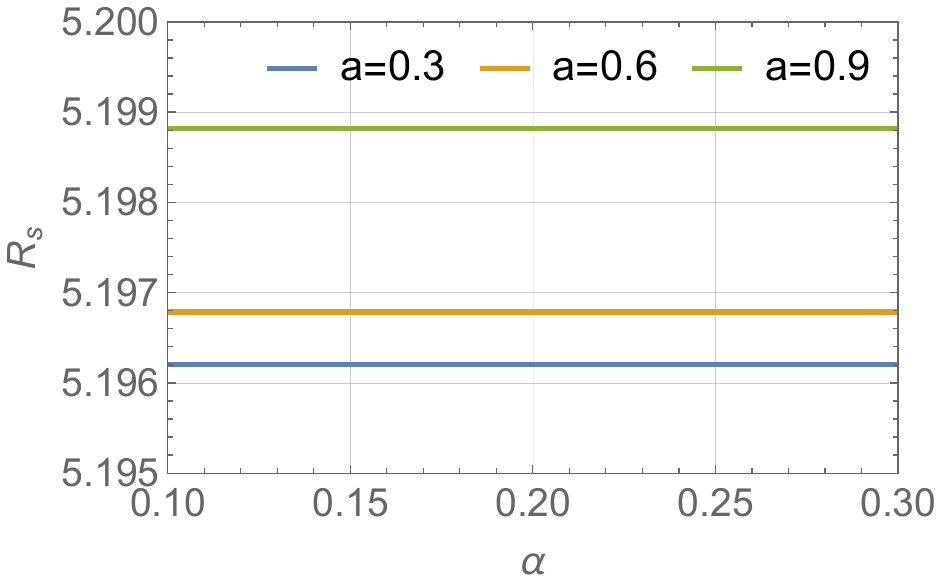}
}
{
\includegraphics[width=0.31\columnwidth]{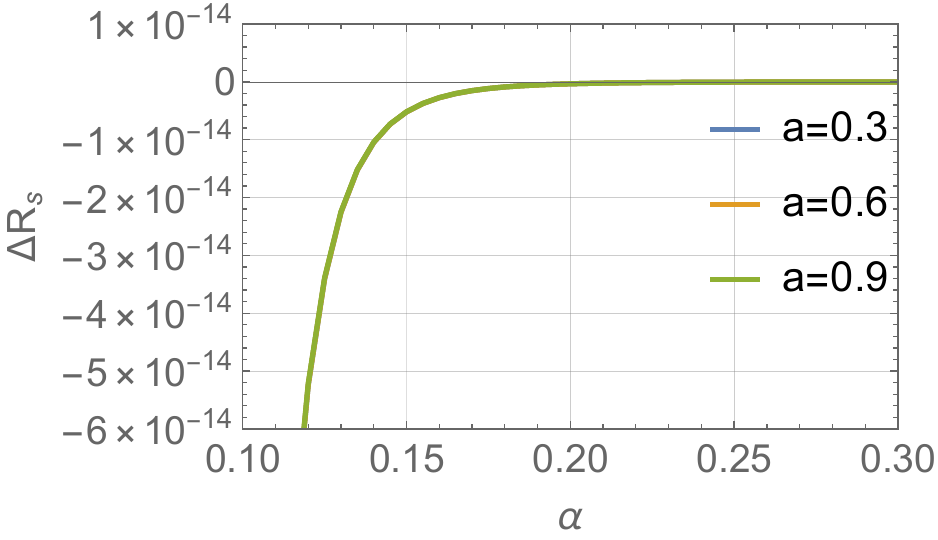}
}
{
\includegraphics[width=0.31\columnwidth]{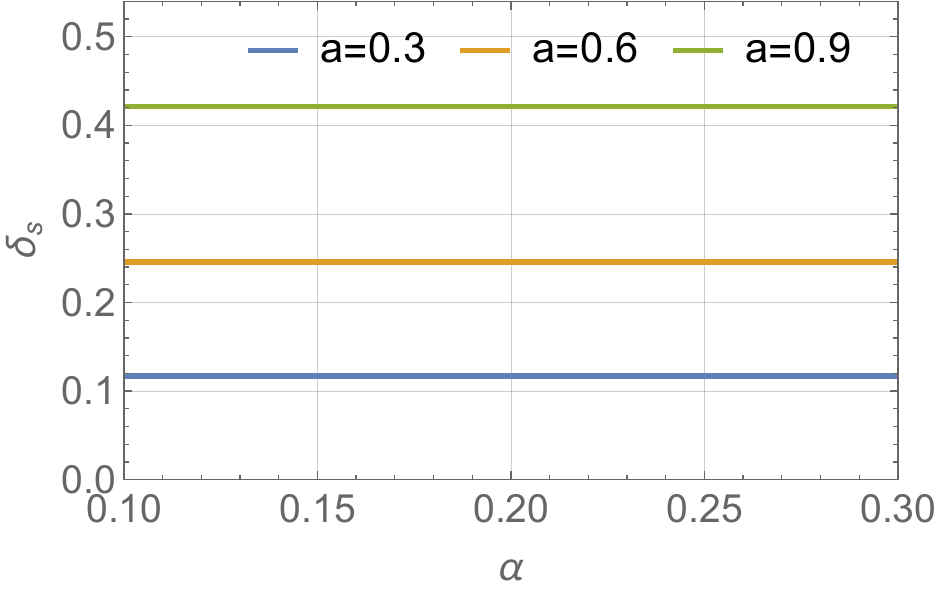}
}
{
\includegraphics[width=0.31\columnwidth]{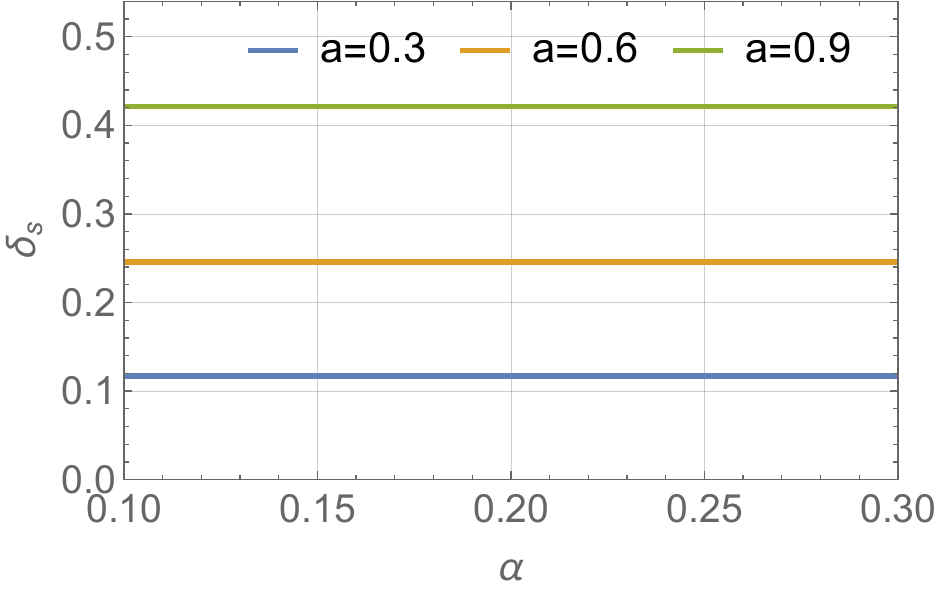}
}
{
\includegraphics[width=0.31\columnwidth]{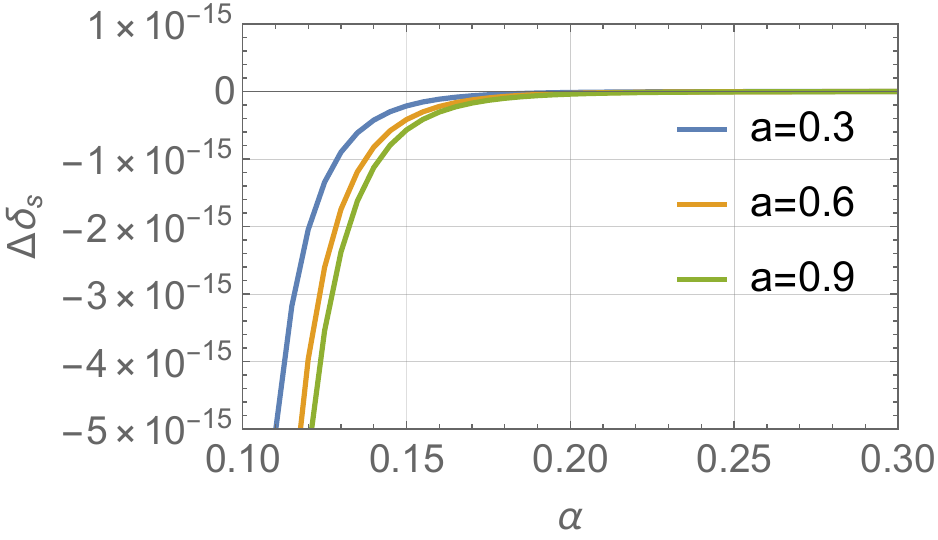}
}
{
\includegraphics[width=0.31\columnwidth]{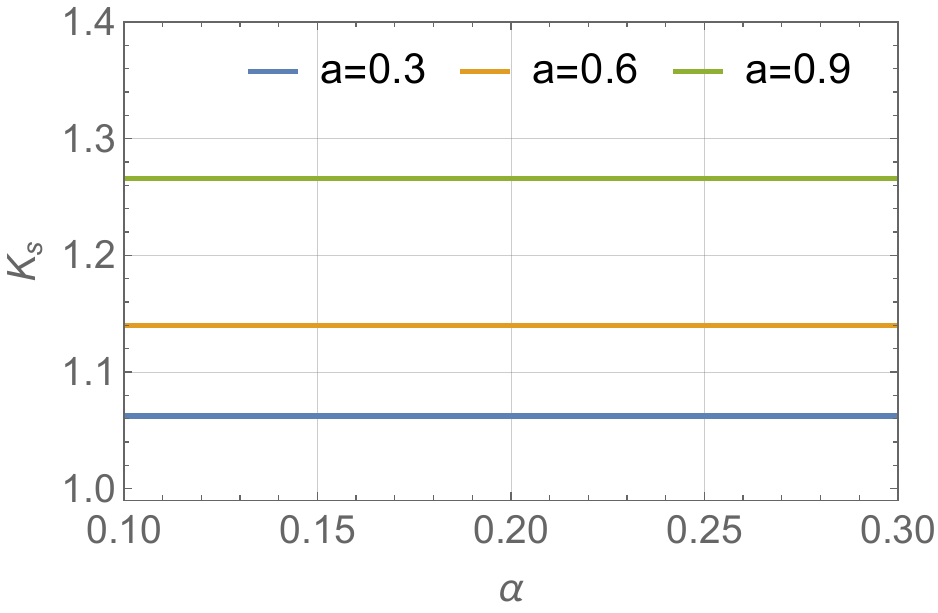}
}
{
\includegraphics[width=0.31\columnwidth]{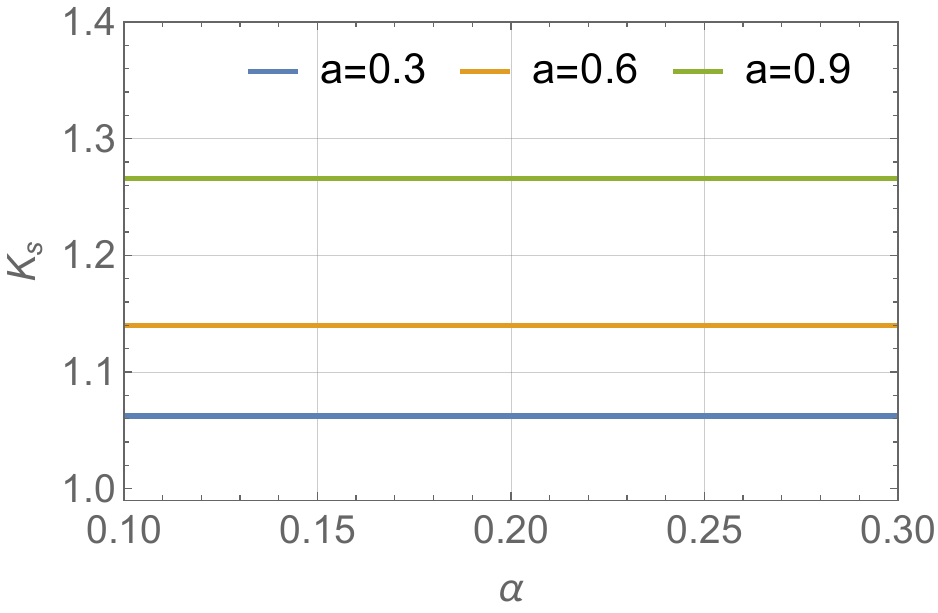}
}
{
\includegraphics[width=0.31\columnwidth]{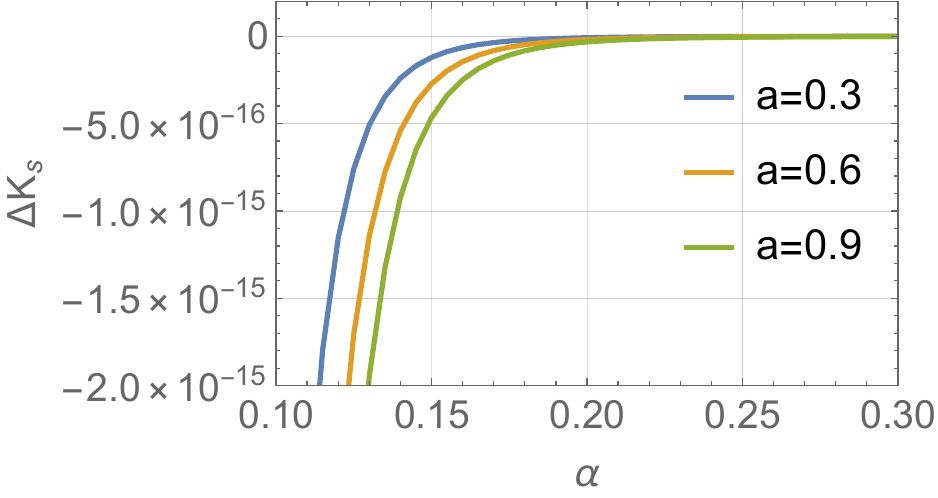}
}
\caption{The observables $R_\text{s}$, $\delta_\text{s}$ and $K_\text{s}$ separately as a function of the shape parameter $\alpha$ of the Kerr spacetime (left panels) and the Kerr-like black hole (middle panels) in the galactic center of M87. The right panels are the differences of these observables between Kerr black hole and Kerr-like black hole. The main calculation parameters are $M=1$, $\rho_\text{e}=6.9 \times 10^{6}$ $M_{\bigodot}/kpc^{3}$, $r_\text{e}=91.2$ $kpc$ and the step size of the shape parameter $\alpha$ is $0.005$ from $0.10$ to $0.30$. We have converted these main calculation parameters to the black hole units before plotting.}
\label{f7}
\end{figure*}

\subsection{The shadow observables of the Kerr-like black hole in a dark matter halo}
In the previous subsection, we introduced the shadow shape of a black hole in a dark matter halo. These black hole shadows have the potential to provide important information about dark matter black holes. Therefore, in this subsection, we start from the observables of black hole shadows and study their impacts on the parameters of the Kerr-like black hole. In other words, we can conversely determine the parameters of the Kerr-like black hole through the shadow shape of the Kerr-like black hole obtained, such as a spin parameter of the Kerr-like black hole. Based on these observables, we may be able to determine or verify the black holes in the universe.

Here, as an example, in Figure  \ref{f5}, we give a coordinate diagram of the Kerr-like black hole shadow ($a=0.999$) in a dark matter halo. The point $C$ corresponds to an unstable circular orbit when the observer looks from the equatorial plane. Therefore, in Figure \ref{f5}, we mainly use the radius of this approximate circle (the black dotted line) to determine the radius $R_s$ of the shadow of the black hole. Besides, we also define the distortion parameter $d_s=\alpha_l-\alpha_l'$, which represents the deformation degree of the left side of the shadow. These two parameters $R_s$ and $d_s$ are the important observables of the black hole shadow. In addition, the authors also studied two other observables, the so-called the area of shadow and the oblateness \cite{Wei:2020ght,Lee:2021sws}. Next, we will study these observables strictly following the procedures in Refs. \cite{Hioki:2009na,Wei:2020ght,Lee:2021sws,Ghosh:2020ece,Tang:2022uwi}. For the convenience of discussion, we introduce the coordinates of some important points in Figure \ref{f5}. They are the top point $(\alpha_t,\alpha_t)$, the right point $(\alpha_r,0)$, the bottom point $ (\alpha_b,\alpha_b)$, the left point $(\alpha_l,0)$ and the left point $(\alpha_l',0)$ of the approximate circle. C is the center of the approximate circle with a radius of $R_s$. The center C is on the x-axis, and its coordinate is $(\alpha_r-R_s,0)$. For the shadow of a Kerr-like black hole, the observables can be defined as the shadow radius $R_s$ and the deformation $\delta_s$ respectively. According to the geometry of shadows, these observations $R_s$ and $\delta_s$ can be written as
\begin{equation}
\begin{aligned}
R_s&=\frac{(\alpha _t-\alpha _r)^2+(\beta _t-0)^2}{2(\alpha_ r-\alpha _t)},
\label{e528}
\end{aligned}
\end{equation}
and
\begin{equation}
\begin{aligned}
\delta _s&=\frac{d_s}{R_s}=\frac{\alpha _l-\alpha _l'}{R_s}.
\label{e529}
\end{aligned}
\end{equation}
Besides, a new observable $K_s$, the ratio of $\Delta \beta$ and $\Delta \alpha$, could be used to fit the data of the M87. This ratio $K_s$ has the following form in shadow geometry
\begin{equation}
\begin{aligned}
K_s=\frac{\Delta \beta}{\Delta \alpha}=\frac{\beta_t-\beta_b}{\alpha_r-\alpha_l}=\frac{2\beta_t}{(2-\delta _s)R_s},
\label{e530}
\end{aligned}
\end{equation}
where, $\alpha_r=2R_s + \alpha_l'$. The definitions (\ref{e528}), (\ref{e529}) and (\ref{e530}) are of great significance that they establish the relationship between the observables and the parameters of a Kerr-like black hole. Therefore, we can determine some important parameters of a black hole based on the shape of its shadow. These coordinates in above equations can be calculated with the help of (\ref{e527}). Next, we first study the impacts of the  observables on the rotation parameter of this Kerr-like black hole in the M87, and compared with the case of the Kerr black hole. In Figure \ref{f6}, we present these three observables separately as a function of the rotation parameter. Our results show that these observables increase with the increasing of the rotation parameter $a$ both in the Kerr spacetime and the Kerr-like black hole. However, the difference of these observables between these two black holes is not obvious. Therefore, we define three observable differences to determine the size relationship between them, that is $\Delta R_\text{s}=R_\text{s}(\text{Kerr})-R_\text{s}(\text{DM})$, $\Delta \delta_\text{s}=\delta_\text{s}(\text{Kerr})-\delta_\text{s}(\text{DM})$ and $\Delta K_\text{s}=K_\text{s}(\text{Kerr})-K_\text{s}(\text{DM})$. Particularly, it can be seen from the observable difference $\Delta R_s$ that the shadow radius of a Kerr-like black hole is slightly larger than that of a Kerr black hole and the difference $\Delta R_\text{s}$ increases with the increasing of the rotation parameter..  The results of the black hole shadow radius $\Delta R_\text{s}$ are consistent with the results in subsection \ref{s51}, which shows that the event horizons of a Kerr-like black hole in a dark matter halo are slightly larger than the Kerr black hole. Similarly, the size between the Kerr-like black hole and the Kerr black hole in the observable differences $\Delta \delta_\text{s}$ and $\Delta K_\text{s}$ are the same as the $\Delta R_\text{s}$ and the values of the differences are negative. These results indicate that these two observables $\Delta \delta_\text{s}$ and $\Delta K_\text{s}$ of the Kerr-like black hole is slightly larger than that of the Kerr black hole. For example, when the rotation parameter is $a=0.8$, the observables of these Kerr-like black holes in the M87 are $R_\text{s} =5.1979167414914803103$, $\delta_\text{s}=0.35114055434480332020$ and $K_\text{s}=1.2125479519509213481$. While the Kerr black hole with the rotation parameter $a=0.8$, these observables are $R_\text{s} =5.1979167414914775730$, $\delta_\text{s}=0.35114055434480303609$ and $K_\text{s}=1.2125479519509211361$. Besides, these two observable differences $\Delta \delta_\text{s}$ and $\Delta K_\text{s}$  decrease with the increasing of the rotation parameter, and then increase with the increasing of the rotation parameter in the nearly extremal case.  Therefore, the observables of a black hole in a dark matter halo are different from those of the Kerr black hole. In other words, with the development of observation technology, if we observe these observables in the galaxy center of M87, it is possible to provide a reliable basis for the existence of the Kerr-like black hole in a dark matter halo.

\indent On the other hand, these observables of the Kerr-like black hole may help to examine the density profile in a dark matter halo. In the density profile of dark matter, the parameter $\alpha$ is the shape parameter of the halo. Therefore, here we study the impacts of observables on the shape parameter $\alpha$ in the galactic center of M87. In Figure \ref{f7}, we present these observables as a function of the shape parameter $\alpha$ both in the Kerr spacetime and the Kerr-like black holes in a dark matter halo. In the left panels, since the Kerr black hole is independent of the parameter $\alpha$, these observables do not change as $\alpha$ increases. In the middle panels of Figure \ref{f7}, we present these observables of the Kerr-like black hole. We find that these observable changes are small with shape parameter, but they all increase with the increasing of the rotation parameter. Therefore, in order distinguish Kerr black hole and Kerr-like black hole, we define these differences ($\Delta R_s, \Delta \delta_s, \Delta K_s$) as discussed above. In the right panels of Figure \ref{f7}, we present the differences between these observables. We find that, for a particular rotation parameter, these observable differences are trend to zero as the shape parameter increases. This shows that as the shape parameter increases, it becomes difficult to distinguish the Kerr black holes from the Kerr-like black holes. However, the black hole in a dark matter halo must be different from the GR black hole. Therefore, there is a theoretical upper limit to the shape parameter from the perspective of the black hole shadow observables. Based on the numerical results of the black hole shadow observable differences in the right panels of Figure \ref{f7}, the upper limit we given is approximately $\alpha<0.22$, and this is equivalent to verifying the rationality of the authors' hypothesis in \cite{Wang:2019ftp} based on the density profile of dark matter from the perspective of the black hole shadow observables. Besides, based on these differences, we find that these observables for Kerr-like black holes are always slightly larger than for Kerr black holes under the same shape and rotation parameter. As the rotation parameter increases, these differences $\Delta \delta_\text{s}$ and $\Delta K_\text{s}$ gradually increase, but the difference $\Delta R_\text{s}$ is not obvious. Similarly, this method can also be used to constrain and test other parameters of the dark matter model, such as the upper limit of the density of dark matter. Since the basis of our research is based on the mass model of M87, the halo density and characteristic radius of dark matter are uniquely determined. Therefore, we only studied the impacts of shape parameters to black hole shadow observables in a dark matter halo. Overall, it is possible that the observables from the black hole shadow provide a new method to examine the density parameter of dark matter. With the increase of observational data, these observables of the black hole shadows may further enrich and improve dark matter density profiles and dark matter models.

\section{Conclusions and Discussions}\label{s6}
\indent In this paper, using the Einasto profile in a dark matter halo, we obtained the metric of the Schwarzschild-like black hole located in the galactic center of M87. And then using the Newman-Janis algorithm, we extended our solution to the case of the Kerr-like black hole, and this solution was satisfied with the Einstein field equations. Besides, we also studied and analyzed the basic physical properties of these black holes in the galactic center of M87 and then compared some results of them with the Kerr black hole. We found that dark matter may have a positive impact on the physical properties of black hole. Our conclusions are as follows:\\
\indent (1) When the black hole is in a dark matter halo, our results show that the dark matter may effectively increase the event horizons of the black hole and expand the range of the ergosphere of the Kerr-like black hole.

\indent (2) In a dark matter halo, for the area of ergosphere of the Kerr-like black hole, that is the outer horizon and outer infinite redshift surface decrease with the increase of rotation parameter $a$, while the inner horizon and inner infinite redshift surface increase with the increase of rotation parameter $a$.

\indent (3) In a dark matter halo, the motion equation for neutral particles near the black hole is given by (\ref{e59}). In particular, for photons in the equatorial plane, the motion equation of the photons is given by Eq. (\ref{e510}).

\indent (4) In a dark matter halo, the geodesic equation for light-like particles is given by Eqs. (\ref{e517}). By calculating the circular orbit of the photon, we find that the photon region $r_\text{c}$ is theoretically between the maximum and minimum values.

\indent (5) In a dark matter halo, for the shadow of the Kerr-like black hole, our results show that with the increasing of the rotation parameter $a$, the degree of distortion of the black hole shadow gradually increases. Besides, we also study the impact of different dark matter parameters on the black hole shadow, and compare them with the Kerr black hole (i.e., $\rho_\text{e}=0$). We find that the shape of the black hole shadow in a dark matter halo is similar to the Kerr black hole. It is difficult to distinguish the Kerr-like black hole and the Kerr black hole from their shapes of black hole shadow while ensuring the integrity of the black hole's shadow. However, since the event horizon of the black hole in the dark matter halo is larger than that of the Kerr black hole, its shadow may be slightly larger than that of the GR black hole.

\indent (6) Based on the geometry of the black hole shadow, we define three observables $(R_\text{s}, \delta_\text{s}, K_\text{s} )$ and establish a connection between the observables and the black hole parameters. With the observables from shadows, we find that these observables increase with the increasing of the rotation parameter both in the Kerr-like black hole and Kerr black hole. We find that the shadow radius $R_s$ of the Kerr-like black holes is slight larger than that of the Kerr black hole, which is consistent with the conclusions of (1) and (5). Similarly, the size between the Kerr-like black hole and the Kerr black hole in the observable differences $\Delta \delta_\text{s}$ and $\Delta K_\text{s}$ are the same as the $\Delta R_\text{s}$ and the values of the differences are negative. These results indicate that these two observables $\Delta \delta_\text{s}$ and $\Delta K_\text{s}$ of the Kerr-like black hole is slightly larger than that of a Kerr black hole. Therefore, from the perspective of the observables, the Kerr-like black hole and the Kerr black holes can be distinguished. On the other hand, we find that these observables of the Kerr-like black hole may help to examine the density profiles in a dark matter halo. From the perspective of these observables and the fact that the Kerr-like black hole and the Kerr black hole is distinguishable, we give an upper limit of the shape parameter of the density profile in a dark matter halo, that is approximately $\alpha<0.22$, which is basically consistent with the assumption the authors used in Ref. \cite{Wang:2019ftp}. Therefore, it is possible that these observables from the black hole shadow provide a new method to examine the density parameter of dark matter.

\indent (7) In the future, these results of the black hole in a dark matter halo may be detected, which indirectly provides an effective method for detecting the existence of dark matter. Meanwhile, with the increase of observational data, these observables of the black hole shadows may further enrich and improve density profiles of dark matter, even the dark matter models.

\indent At last, it would be interesting to study the quasinormal modes of a black hole in a dark matter halo. The quasinormal mode is the main mode after the merger of binary black holes, and it will carry important information of the black hole. Meanwhile, the black hole shadows and photon rings can also be used to detect the existence of the black hole. Therefore, in the next step of work, we will study the quasinormal mode of the black hole and the photon ring phenomenon gradually, expecting to provide some directions for the existence of the black hole in a dark matter halo.

\appendix
\section{Einstein field equations}\label{sa}
In this appendix, based on the metric (\ref{e412}), we will prove that this metric is a solution to Einstein field equations. Firstly, we rewrite (\ref{e412}) in the Kerr-like form
\begin{equation}
\begin{aligned}
ds^2= & -(1-\frac{2R(r)r}{\Sigma }) dt^2 +\frac{\Sigma}{\Delta } dr^2 -\frac{4 R(r) r a \sin ^2 \theta }{\Sigma } dt d \phi + \Sigma d\theta^2 \\
&+\left ( (a^2+r^2)\sin^2 \theta +\frac{2 R(r) r a^2\sin^4\theta }{\Sigma }\right ) d\phi^2,
\label{ea1}
\end{aligned}
\end{equation}
where,
\begin{equation}
\begin{aligned}
2R(r)=2M + r -r g(r), \Delta =r^2-2R(r)r+a^2, \Sigma=r^2+a^2\cos^2 \theta.
\label{ea2}
\end{aligned}
\end{equation}
Considering that the Kerr-like black hole in a dark matter halo, it should satisfy the Einstein field equations, that is
\begin{equation}
\begin{aligned}
G_{\mu\nu}=R_{\mu\nu}-\frac{1}{2}g_{\mu\nu}R=8\pi T_{\mu\nu},
\label{ea3}
\end{aligned}
\end{equation}
Taking the metric (\ref{ea1}) into the Einstein field equations, we can obtain the following non-zero Einstein tensors
\begin{equation}
\begin{aligned}
&G_{tt}=  \frac{2(a^2r^2+r^4-a^4\cos^2\theta \sin^2\theta -2r^3R(r) )R'(r)}{\Sigma ^3} -\frac{a^2r\sin^2\theta R''(r)}{\Sigma ^2},      \\
&G_{r r} =   -\frac{2r^2R'(r)}{\Sigma \Delta },     \\
&G_{t\phi}=   -\frac{2a\sin^2 \theta ((r^a+a^2)(a^2\cos^2\theta-r^2)+2r^3R(r))R'(r) }{\Sigma ^3}     +\frac{ar(a^2+r^2)\sin^2\theta R''(r)}{\Sigma ^2}, \\
&G_{\theta \theta }=  -\frac{2a\cos^2\theta R'(r)}{\Sigma } -rR''(r)      \\
&G_{\phi \phi }= -\frac{a^2\sin^2\theta((a^2+r^2)(a^2+(a^2+2r^2)\cos 2\theta)+4r^3\sin^2 \theta R(r))R'(r)}{\Sigma ^3} \\
&-\frac{r\sin^2\theta (a^2+r^2)^2R''(r)}{\Sigma ^2}.
\label{ea4}
\end{aligned}
\end{equation}
On the other hand, in Eq. (\ref{ea3}), $T_{\mu\nu}$ is the energy momentum tensor and $T^{\mu\nu} =e^\mu_a e^\nu_b T^{ab}$, and $T^{ab}=diag[1/(-\rho_\epsilon, p_r, p_\theta,p_\phi)]$. Then the inverse of the energy-momentum tensor can be represented in the orthonormal basis as follows
\begin{equation}
\begin{aligned}
T^{\mu \nu}=-1/\rho_\epsilon e_t^\mu e_t^\nu + 1/p_r e_r^\mu e_r^\nu + 1/p_\theta e_\theta^\mu e_\theta^\nu + 1/p_\phi e_\phi^\mu e_\phi^\nu,
\label{ea5}
\end{aligned}
\end{equation}
where, the forms of the orthonormal bases read
\begin{equation}
\begin{aligned}
&e^\mu_t=\frac{1}{\sqrt{\Sigma \Delta} }(r^2+a^2,0,0,a),\\
&e^\mu_r=\sqrt{\frac{\Delta }{\Sigma}}(0,1,0,0),\\
&e^\mu_\theta=\frac{1}{\sqrt{\Sigma }}(0,0,1,0),\\
&e^\mu_\phi =-\frac{1}{\sqrt{\Sigma \sin^2\theta }}(a \sin^2 \theta,0,0,1),\\
\label{ea6}
\end{aligned}
\end{equation}
Therefore, taking Eqs. (\ref{ea3}), (\ref{ea4}) and (\ref{ea6}) into account, we can obtain
\begin{equation}
\begin{aligned}
&8\pi \rho_\epsilon =-e^\mu_t e^\nu_t G_{\mu\nu},\\
&8\pi p_r =e^\mu_r e^\nu_r G_{\mu\nu}=g^{rr}G_{rr},\\
&8\pi p_\theta  =e^\mu_\theta  e^\nu_\theta  G_{\mu\nu}=g^{\theta \theta }G_{\theta \theta },\\
&8\pi p_\phi =-e^\mu_\phi e^\nu_\phi G_{\mu\nu},\\
\label{ea7}
\end{aligned}
\end{equation}
Finally, we obtain expressions for the energy momentum tensors of a rotating black hole
\begin{equation}
\begin{aligned}
\rho_\epsilon =-p_r=\frac{2r^2R'(r)}{8\pi\Sigma ^2}, p_\theta=p_\phi=p_r-\frac{rR''(r)+2R'(r)}{8\pi\Sigma },
\label{ea8}
\end{aligned}
\end{equation}
where, $R'(r)$ denotes $dR(r)/dr$. These results indicate that (\ref{e412}) is indeed a solution to Einstein field equations.

\acknowledgments
We would like to acknowledge the anonymous referees for the constructive reports, which significantly improved this paper. This research was funded by the National Natural Science Foundation of China (Grant No.12265007) and the Natural Science Special Research Foundation of Guizhou University (Grant No.X2020068).


\bibliographystyle{epjc}
\bibliography{epjcexample}
\end{document}